\documentclass[aps, preprint, groupedaddress, floatfix]{revtex4}
\usepackage{epsfig}
\usepackage{graphicx}
\usepackage{amsmath}
\usepackage{bbold}
\usepackage{tabularx, booktabs}
\usepackage[normalem]{ulem}
\usepackage{soul}  
\usepackage{xcolor} 
\usepackage{multirow} 
\usepackage{xr}

\newcommand{\ie}{{\textit i.e.}}
\newcommand{\eg}{{\textit e.g.}}
\newcommand{\etal}{{\textit et al.}}
\newcolumntype{Y}{>{\centering\arraybackslash}X}
\newcolumntype{Z}{>{\hsize=1.1\hsize\centering\arraybackslash}X}

\newcommand*{\citen}[1]{%
  \begingroup
    \romannumeral-`\x 
    \setcitestyle{numbers}%
    \cite{#1}%
  \endgroup
}

\begin{document}

\title{Two-dimensional ferroelectric tunnel junction:
  the case of monolayer In:SnSe/SnSe/Sb:SnSe homostructure
  }

\author{Xin-Wei Shen$^{1}$, Yue-Wen Fang$^{1,2*}$, Bo-Bo Tian$^{1}$, Chun-Gang Duan$^{1,3}$}
\email{fyuewen@gmail.com (Y.-W.F); cgduan@clpm.ecnu.edu.cn (C.-G.D)}
\affiliation{
$^{1}$State Key Laboratory of Precision Spectroscopy and Key Laboratory of Polar Materials and Devices, Ministry of Education, Department of Optoelectronics, East China Normal University, Shanghai, 200241, China \\
$^{2}$Department of Materials Science and Engineering, Kyoto University, Kyoto 606-8501, Japan \\
$^{3}$Collaborative Innovation Center of Extreme Optics, Shanxi University, Taiyuan, Shanxi 030006, China \\
}

\keywords{\textit{2D materials, ferroelectrics, electron transport, TER effect, memory device}}

\begin{abstract}
    Ferroelectric tunnel junctions, in which ferroelectric polarization
    and quantum tunneling are closely coupled to induce
    the tunneling electroresistance (TER) effect, have
    attracted considerable interest due to their
    potential in non-volatile and low-power consumption
    memory devices. The ferroelectric size effect, however, has
    hindered ferroelectric tunnel junctions from exhibiting
    robust TER effect.
    Here, our study proposes doping engineering in a two-dimensional
    in-plane ferroelectric semiconductor
    as an effective
    strategy to design a two-dimensional ferroelectric tunnel junction composed of
    homostructural $p$-type semiconductor/ferroelectric/$n$-type semiconductor.
    Since the in-plane polarization persists in the monolayer ferroelectric barrier, the vertical thickness of
two-dimensional ferroelectric tunnel junction can be as thin as monolayer.
    We show that the monolayer In:SnSe/SnSe/Sb:SnSe junction provides
    an embodiment of this strategy.
    Combining density functional theory calculations with
     non-equilibrium Green's function formalism, we investigate
    the electron transport properties of In:SnSe/SnSe/Sb:SnSe and
    reveal a giant TER effect of 1460$\%$.
    The dynamical modulation of
    both barrier width and barrier height during the ferroelectric switching
    are responsible for this giant TER effect.
    These findings provide an important insight towards the understanding
    of the quantum behaviors of electrons in materials at the two-dimensional limit,
    and enable new possibilities for next-generation non-volatile
    memory devices based on flexible two-dimensional lateral ferroelectric
    tunnel junctions.
\end{abstract}
\maketitle

\vspace{1cm}
\section*{INTRODUCTION}
The past decades have witnessed an explosion in the field of
ferroelectric materials~\cite{Kalinin_2010review,bowen2014review,martin2017NRMater},
headlined by the design
of ferroelectric tunnel junctions (FTJs) with the aim of
accelerating their commercial applications into
non-volatile information devices
~\cite{PhysRevLett.94.246802,tsymbal2006ey,PhysRevLett.98.137201,scott2007NatMater,garcia2009giant,garcia2010Science,chanthbouala2012NatMater,LuAPR-2015review,hu2016NatComm,boyn2017learning,huang2018solid}.
FTJs are composed of two metallic electrodes separated by a thin
ferroelectric barrier.
The information is encoded via the non-volatile
ferroelectric polarization that can be electrically switched.
Switching the polarization gives rise to dramatic change of the
tunneling electroresistance (\ie TER effect~\cite{PhysRevLett.94.246802}), making it possible to non-destructively
readout the polarization state that carries information.
The increasing suppression of ferroelectricity by the depolarization field
as ferroelectric materials are reduced down to nanometers
~\cite{Ghosez2003Nature,spaldin2004fundamental},
or in other words, the ferroelectric size effect~\cite{Li_1997JSAP}, however,
has impeded the development of nanometer-size FTJs.
Although several widely studied ferroelectric oxides (\eg, BaTiO$_3$)
have been experimentally proved to show switchable polarization
down to the thickness of
1-4 unit cells~\cite{fong2004ferroelectricity,tenne2006science,Tenne-PRL2009,Ramesh2012-PRB},
the
lack of stability and reproducibility
at room temperature of FTJs
based on these conventional ferroelectric thin films
cannot keep up with the ever-growing commercial needs for the ultra-low-power, high-speed,
and non-volatile nanoscale memory devices.

In spite of the difficulties in growing thin films
of ferroelectric oxides due to their high demand
for the growth conditions~\cite{choi2004enhancement,PhysRevLett.96.127601,lu2012enhancement},
some early first-principles studies have predicted the existence of in-plane ferroelectricity
in monolayer binary inorganic compounds~\cite{PhysRevLett.112.157601,PhysRevB.91.161401}.
These works have prompted the active search for the two-dimensional (2D)
ferroelectrics both in theory and experiment
~\cite{chang2016discovery,Fei2016PRL,WuMengHao-nanolett-2016,
MHWu2016NanoLett-2D-FE,TingHu-NanoLett2016,Haleoot-PRL2017,
XFQian-2DMater-2017,ding2017NC,CXHuang-PRL2018,Poh-MBE-In2Se3-nanolett2018,
PhysRevLett.120.227601,Cui-NanoLett2018,JLiu-2DMater-2019}.
Among these studies, the group-IV chalcogenide semiconductors have shown
to be one class of the most promising 2D ferroelectric materials.
As reported in the experimental study by Chang \etal,
atomic-thick SnTe exhibits near-room-temperature in-plane ferroelectricity,
and the ferroelectricity in ultrathin SnTe films above 2 unit cells can persist
above room temperature~\cite{chang2016discovery}.
A nonvolatile in-plane ferroelectric random access memory (FeRAM)
based on 3 unit cells SnTe is designed in Chang~\etal's study
to readout information non-destructively,
which is more superior than
conventional FeRAM where reading is destructive~\cite{chang2016discovery}.
In addition to the 2D monochalcogenide SnTe, the 2D trichalcogenide
$\alpha$-In$_2$Se$_3$, in which
out-of-plane and in-plane polarization are intrinsically intercorrelated,
can sustain the ferroelectricity
up to 700 K~\cite{PhysRevLett.120.227601}.
By integrating 2D $\alpha$-In$_2$Se$_3$
into a ferroelectric Schottky diode junction,
high current density of $\sim$~12 A/cm$^{2}$ is reported to be more than
two times of the conventional ferroelectric diode junctions~\cite{Poh-MBE-In2Se3-nanolett2018}.
The high-quality monolayer SnTe and $\alpha$-In$_2$Se$_3$\ can
be prepared by molecular beam epitaxial technique or physical vapor deposition~\cite{chang2016discovery,Zhou-PVD-NanoLett-In2Se3-2015,Poh-MBE-In2Se3-nanolett2018},
indicating the feasibility of synthesis of 2D ferroelectric materials.
These recent successes of maintaining stable electric polarization in 2D semiconductors,
which are both from an experimental as well as from a theoretical perspective,
motivate us to explore their interesting properties and potential
applications in
ferroelectric non-volatile memories based on 2D-FTJs.

Herein, through doping engineering in 2D semiconductors to establish electrodes (\ie, $p$-type
and $n$-type semiconductors),
in combination with the coupling between the robust in-plane ferroelectricity and quantum
tunneling in ferroelectric semiconducting barrier at the two-dimensional limit,
we theoretically design a new class 2D-FTJs based on homostructure.
Using a model which takes
into account screening of polarization charges in electrodes,
charge accumulation/depletion at semiconductor/ferroelectric interfaces,
reversible metallization of the ferroelectric barrier~\cite{PhysRevLett.116.197602},
and direct quantum tunneling across a
ferroelectric barrier, we demonstrate the possibility to obtain
giant TER effect in 2D-FTJ.
As an example 2D-FTJ, In:SnSe/SnSe/Sb:SnSe homostructure
is investigated by
first-principles calculations with non-equilibrium Green's function formalism.
We find a giant TER effect of 1460$\%$, which is
on account of a dual modulation of both barrier width
and barrier height.
To our knowledge, this is the first demonstration of the 2D-FTJ
based on homostructure that holds promise in future memory devices.

The geometry optimizations and electronic structure calculations of
slab models are performed within density-functional theory (DFT)
using the projector augmented wave (PAW) method~\cite{Blochl1994}, as
implemented in the Vienna ab initio Simulation Package
(VASP)~\cite{Kresse-PRB-1996,PhysRevB.59.1758}. The exchange correlation functional
is treated in
generalized gradient approximation (GGA) with the type of Perdew-Burke-Ernzerhof (PBE)~\cite{PhysRevLett.77.3865}.
The kinetic-energy cutoff of 500 eV is applied to the plane wave expansion and a
$\Gamma$-centered 1$\times$12$\times$1 \textbf{k} points grid is adopted for Brillouin zone sampling.
All the structures are optimized until the Hellmann-Feynman forces are below
1 meV/\AA~, and the convergence threshold of electronic energy is 10$^{-6}$ eV.
A vacuum space of 15~\AA~is used to avoid interactions between adjacent layers.
The spontaneous ferroelectric polarization ($P_s$) is determined by the
Berry phase method~\cite{PhysRevB.47.1651,RevModPhys.66.899}.

The device properties of the 2D-FTJ are calculated using density functional
theory plus non-equilibrium Green's function formalism (DFT+NEGF approach)
~\cite{PhysRevB.63.245407,PhysRevB.65.165401}
 as implemented in Atomistix ToolKit-Virtual NanoLab (ATK-VNL) software package
~\cite{QuantumWise-website}.
Double-$\zeta$ plus polarization basis set is employed, and a real-space mesh cut-off
energy of 80 Hartree is used to guarantee the good convergence of the device configuration.
The electron temperature is set at 300 K. The 1$\times$21$\times$101 \textbf{k} mesh is used for the
self-consistent calculations to eliminate the mismatch of Fermi level between electrodes and
the central region. An increased number 201 of \textbf{k} points along $y$ axis is adopted during the
calculations of transmission spectra and spatially resolved device density of states.

\section*{RESULTS AND DISCUSSION}
\begin{figure}[!t]
\includegraphics[angle=0,width=1\textwidth]{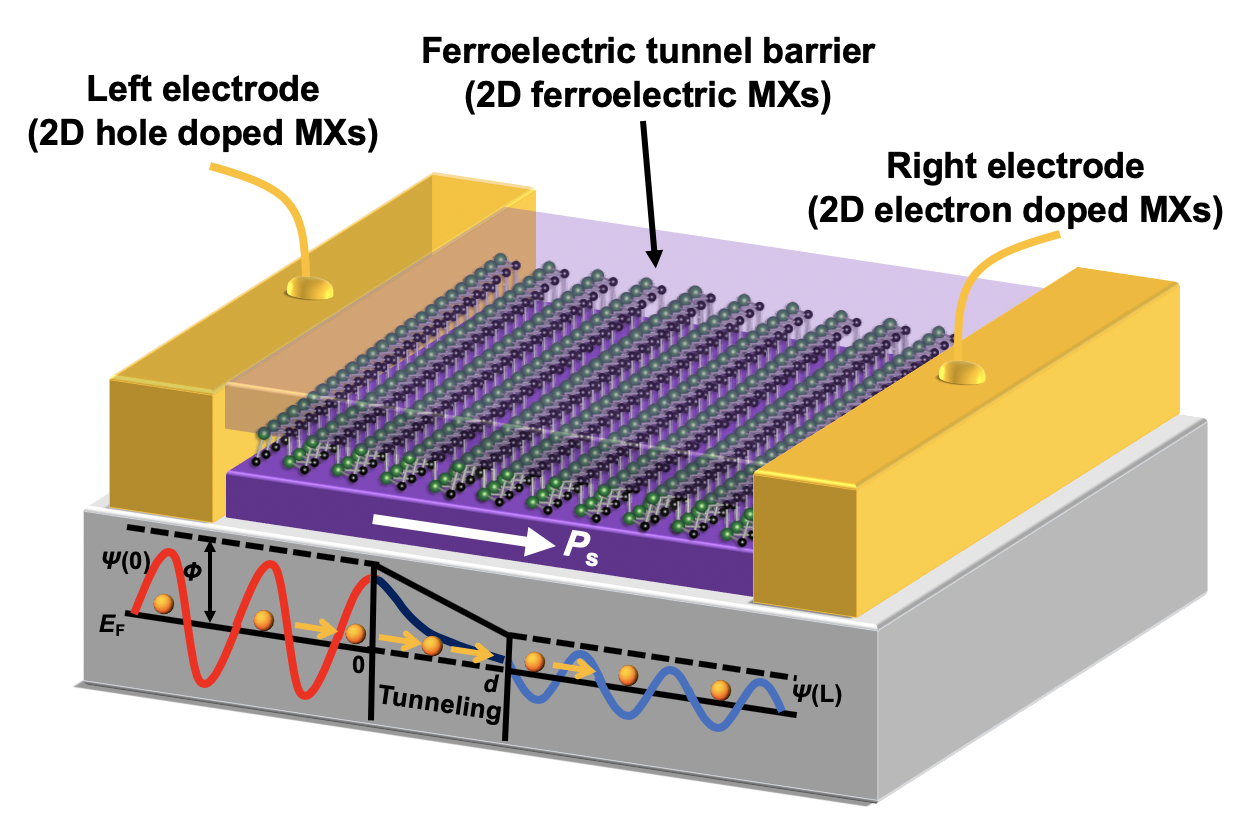}
\caption{\label{fig:Fig1}
Schematic diagram of a two-dimensional ferroelectric tunnel junction (2D-FTJ) device based on
homostructure.
A pure 2D monochalcogenide (MX) semiconductor with in-plane ferroelectricity serves as a tunneling barrier.
The left and right electrodes are obtained by $p$-type and $n$-type doping of the same MX.
The inset illustrates the wave-particle duality
of the quantum tunneling and the change of barrier $\mathit{\Phi}$ in response to the polarization reversal.
Governed by quantum mechanics,
the electrons tunnel across the barrier in the form of
evanescent state which decays exponentially through the barrier in amplitude.
}
\end{figure}

Different from the architecture of conventional vertical FTJs
built on heterostructures~\cite{garcia2014ferroelectric},
the 2D-FTJ in our study makes the utmost of the structural features
of 2D ferroelectrics.
We take 2D ferroelectric group-IV monochalcogenide
(\eg, SnSe~\cite{Fei2016PRL} or SnTe~\cite{chang2016discovery})
as an example to elucidate our model device.
As illustrated in Figure~\ref{fig:Fig1},
a pure 2D ferroelectric monochalcogenide semiconductor
is used as the ferroelectric barrier, on the other hand,
its hole and electron doped forms
(\ie, $p$-type and $n$-type semiconductors)
work as the respective left and right electrodes concurrently.
Hence, this new class of 2D-FTJs
$p$-semiconductor/ferroelectric/$n$-semiconductor ($p$-SC/FE/$n$-SC) are
architected on homostructures in avoid of laminating
several different materials into heterostructures as the conventional FTJs,
which can be expected to reduce the difficulty in device manufacture.

%

In Thomas-Fermi theory, the screening length of a metallic material
is defined as
\begin{equation}
        \label{Eq:screen-length}
\delta = \frac{1}{e} \sqrt{\frac{\varepsilon}{\rho}}
\end{equation}
where $\varepsilon$ is the dielectric permittivity and $\rho$
is the density of states at the Fermi level ${E_{\rm F}}$.
In the homostructural 2D-FTJ displayed in Figure~\ref{fig:Fig1},
the dielectric permittivity of
left/right electrode is equivalent to that of
pure 2D ferroelectric group-IV monochalcogenide
at saturation polarization~\cite{PhysRevLett.109.247601}.
Hence, the screening lengths of the two electrodes
only depend on their densities of states at the ${E_{\rm F}}$ that
can be easily controlled by doping.
In order to introduce our model easily,
we assume ${\delta_1 < \delta_2}$, in which
${\delta_1}$ and ${\delta_2}$ are
the screening lengths of left and right electrodes,
respectively.
The model in which ${\delta_1 > \delta_2}$ can also be found
in Supporting Information.

\begin{figure}[!t]
\includegraphics[angle=0,width=0.7\textwidth]{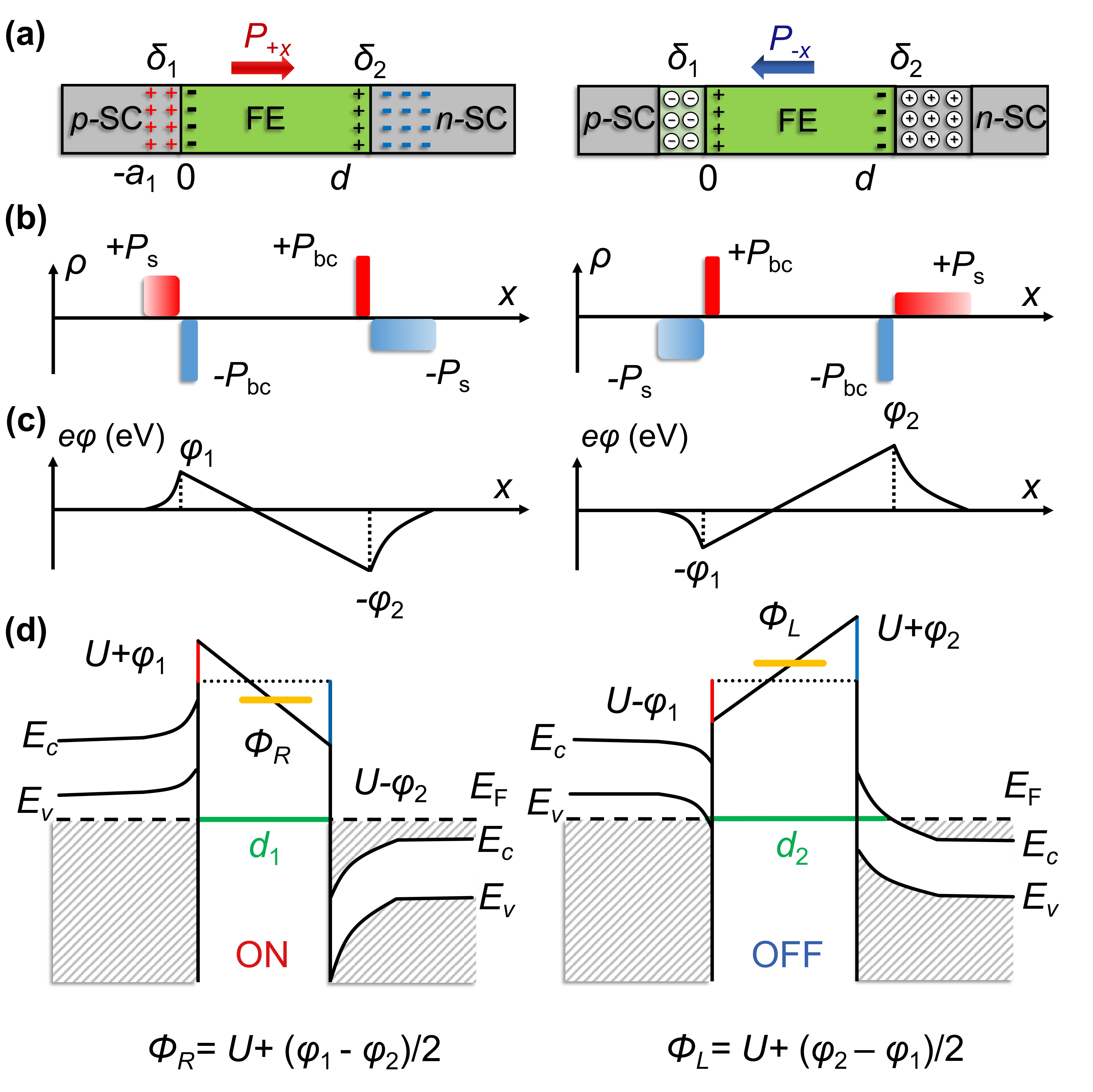}
\caption{\label{fig:Fig2}
Switching mechanisms of the 2D-FTJ $p$-SC/FE/$n$-SC.
(a) The respective schematics of 2D-FTJ in ${P_{+x}}$ (left panel) and ${P_{-x}}$ (right panel) states.
The black ``+'' and ``--'' symbols in the ferroelectric barrier region represent positive and negative ferroelectric
bound charges, respectively.
The red ``+'' in $p$-SC and blue ``--'' in $n$-SC electrodes represent hole and electrons, respectively.
The ``\textcircled{+}'' and ``\textcircled{--}''
represent ionized donors and acceptors respectively.
(b) The distributions of charge densities.
(c) Electrostatic potential energy profiles.
(d) The overall potential energy profiles with corresponding band diagrams.
The barrier width in either ${P_{+x}}$ or ${P_{-x}}$ state is given by the
green line.
The average potential barrier height is indicated by the short orange line.
The Fermi level, as shown by black dashed line, is set as the reference of the barrier height.
}
\end{figure}

Figure~\ref{fig:Fig2}(a) and (b)
schematically depict
the charge distributions and charge density profiles in the 2D-FTJ, where
$d$ is the original width of 2D ferroelectric barrier.
When polarization is rightward (referred to as $P_{+x}$ state), negative bound charge (${-P_{\rm bc}}$)
and positive ferroelectric bound charges (${+P_{\rm bc}}$) are introduced at the left edge and right edge of the
ferroelectric barrier, respectively.
In this case, the hole (electron) carriers in the electrode
of $p$-SC ($n$-SC) \ bb{become} accumulated owning to the spontaneous electric field at the interface (\ie,
ferroelectric polarization field effect).
On the contrary, the carriers in both $p$-SC and $n$-SC electrodes are depleted as
the polarization is leftward (referred to as $P_{-x}$ state).
In order to investigate the change of barrier in response to the ferroelectric switching quantitatively,
we adopt a Thomas-Fermi screening mode~\cite{PhysRevLett.94.246802}.
In this way,
the electrostatic
potentials within left and right electrodes can be written as:
\begin{equation}
        \label{Eq:ThomasModel}
    \varphi_i = \pm \frac{P_s d \delta_i}{\varepsilon_0 \varepsilon_i \left[\varepsilon_{\rm FE}\left(
        \frac{\delta_1}{\varepsilon_1} + \frac{\delta_2}{\varepsilon_2}
        \right) + d \right]}, i = 1~{\rm or}~2 \,,
\end{equation}
where $i$ = 1 for left electrode $p$-SC, $i$ = 2 for right electrode $n$-SC,
$\varepsilon_0$ is the permittivity of free space,
$\varepsilon_{\rm FE}$ is the relative permittivity of the ferroelectric layer,
and ${\varepsilon_i}$ is the dielectric permittivity.
The sign `+' (`--') corresponds to the polarization pointing to (away from) the
studied electrode.
Using Equation~\ref{Eq:ThomasModel} and the
 assumption of ${\delta_1 < \delta_2}$,
we can compare the electrostatic potential
energy at $p$-SC/FE and FE/$n$-SC interfaces:
\begin{equation}
        \label{Ineq:screen}
    |e \varphi_1| \equiv |e \varphi(0)| < |e \varphi_2| \equiv |e \varphi(d)| \,.
\end{equation}
We set the charge of $e$ to `-1' in Equation~\ref{Ineq:screen}
for the simplification of model analysis, it becomes ${|\varphi_1| < |\varphi_2|}$.
This leads to an asymmetry in the electrostatic potential energy profiles
for the opposite
polarization directions, which can be seen in Figure~\ref{fig:Fig2}(c).
In addition to the electrostatic potential energy,
we note that the tunneling electrons should
also overcome electronic potential energy and potential energy barrier inside
2D ferroelectric material.
As pointed out by the earlier studies~\cite{PhysRevLett.94.246802,wen2013ferroelectric}
the potential barriers
can be assumed to be
a rectangular shape of height $U$
with respect to the Fermi level ${E_{\rm F}}$.
Thus, the average barrier height in either
$P_{+x}$ or $P_{-x}$ state can be given as
\begin{eqnarray}
    \begin{cases}
\label{Eq:overallbarrier}
\mathit{\Phi}_R = U + (\varphi_1 - \varphi_2)/2 \\
\mathit{\Phi}_L = U + (\varphi_2 - \varphi_1)/2 \,,
   \end{cases}
\end{eqnarray}
where subscripts $R$ and $L$ correspond to
$P_{+x}$ and $P_{-x}$ states, respectively.
Setting Equation~\ref{Ineq:screen} into Equation~\ref{Eq:overallbarrier}
yields
\begin{eqnarray}
\label{Ieq:overallbarrier_compare}
\mathit{\Phi}_R < \mathit{\Phi}_L \,,
\end{eqnarray}
indicating the average barrier height in $P_{-x}$ state is
higher than that in $P_{+x}$ state. The resulted barrier heights in these two states
are comparatively illustrated
in Figure~\ref{fig:Fig2}(d).

Besides it follows from Figure~\ref{fig:Fig2}(d) that
the effective barrier width can also be changed by the polarization
switching.
As pointed out by some recent
experimental and theoretical
studies~\cite{ZhaoHJ-PhysRevB.97.054107,PhysRevB.99.064106,gu2017coexistence,PhysRevLett.104.147602},
off-center displacements and metallic conductivity
can coexist in doped ferroelectrics.
In our 2D-FTJ model, we consider such a condition where the introduced dopants
in the electrodes
will not eliminate the polar distortions completely,
\ie,
the electrodes enter polar metallic states.
In addition, we note that
charge leakage can be induced at the interface
when a polar metallic material is interfaced with a
ferroelectric material~\cite{fang2019electric,PhysRevLett.116.197602}.
For the $P_{+x}$ state of 2D-FTJ,
carriers are accumulated
at the $p$-SC/FE and $n$-SC/FE interfaces.
This leads to the
enhancement of the charge leakage,
and therefore
the regions of ferroelectric barrier near the two interfaces may
become conductive, resulting in reduction of
the effective tunneling width.
We refer this decreased tunnelling width to as $d_1$ which is smaller than
the original width $d$ of the ferroelectric barrier, as can be seen from
the left panel of Figure~\ref{fig:Fig2}(d).
By contrast, carriers are depleted at the two interfaces in $P_{-x}$ state,
The majority carriers will be partially cancelled out
near the semiconductor/ferroelectric interfaces,
only leaving the immobile ionized acceptors and donors~\cite{wen2013ferroelectric},
as shown in the right
panel of Figure~\ref{fig:Fig2}(a).
Under this condition, charge leakage of $P_{-x}$ state is weaker than
$P_{+x}$ state. As a result, some ferroelectric barrier regions near interfaces will
transform back into insulating state, and the
band alignment in $P_{-x}$ state is also changed with respect to $P_{+x}$ state.
If the charge leakage is sufficiently weak,
all the ferroelectric barrier regions and even partial regions
in the electrodes may become insulating~\cite{PhysRevLett.116.197602}.
In this case, the effective barrier width is increased to $d_2$, which
is larger than effective barrier width $d_1$ of the $P_{+x}$ state.
The modulation of effective barrier width is also known as the
``reversible metallization of the barrier'' as reported by Liu \etal~\cite{PhysRevLett.116.197602}.
It is due to the ferroelectric dual modulation of both barrier width and barrier height
in 2D-FTJ, the tunneling electrons in the $P_{-x}$ state have to overcome additional barrier height
and barrier width. Since the tunneling conductance is determined by the barrier height and barrier
width~\cite{PhysRevLett.94.246802}, the conductance of $P_{+x}$ state will
be dramatically enhanced compared to $P_{-x}$ state.
The higher conductance state is conventionally referred to as ``ON'' state,
with a comparison of the lower
conductance state referred to as ``OFF'' state, as displayed in
Figure~\ref{fig:Fig2}(d).
We note that in conventional FTJs like
Metal-1/ferroelectric/Metal-2 heterostructures~\cite{PhysRevLett.98.137201},
only barrier height is generally modified by polarization switching.
Therefore, the 2D-FTJs in our study can be expected to realize enhanced TER effect.

\begin{figure}[!t]
\includegraphics[angle=0,width=0.9\textwidth]{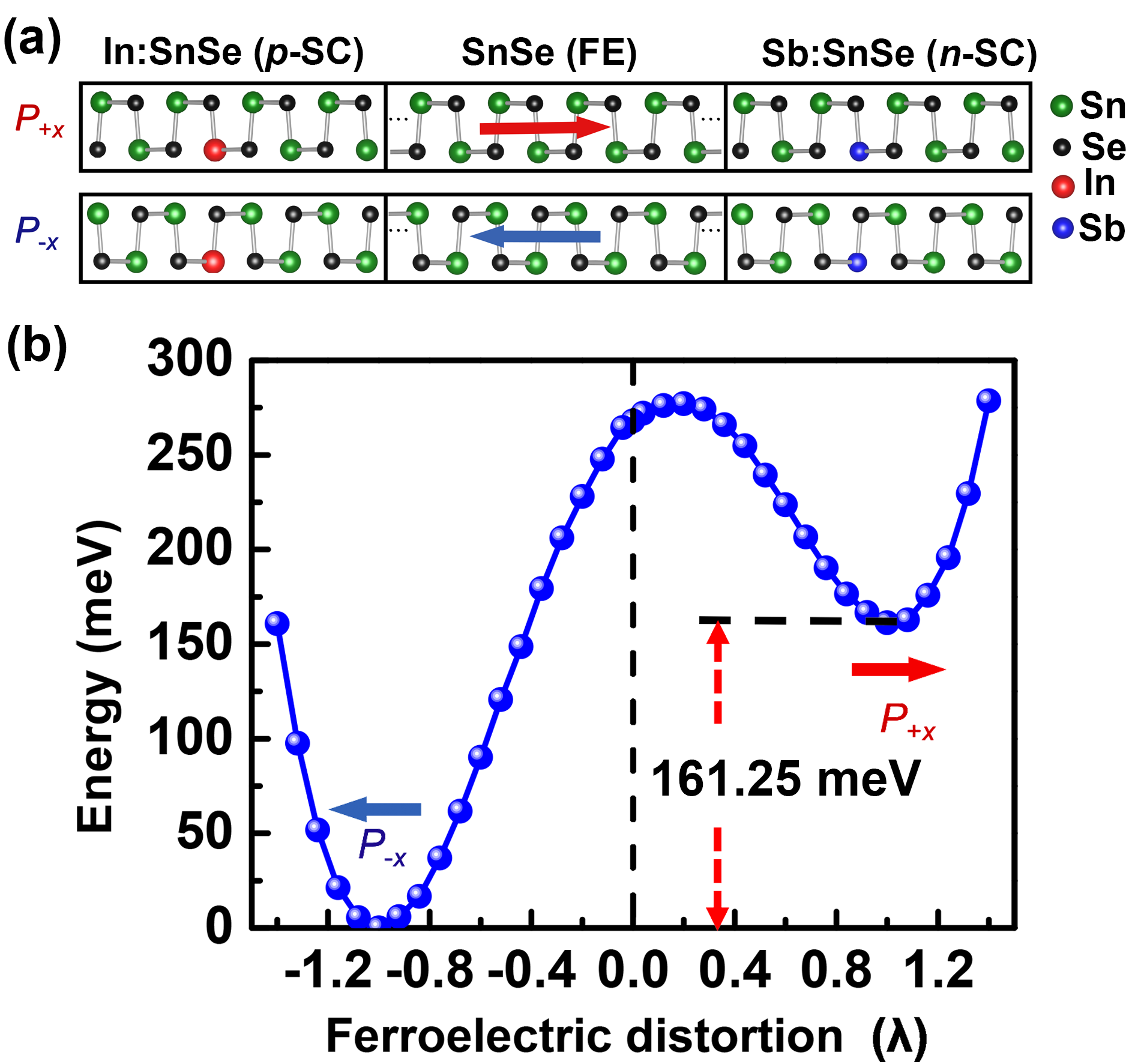}
\caption{\label{fig:Fig3}
(a) The calculated atomic structure and the schematic diagram (shaded regions) of
2D-FTJ In:SnSe/SnSe/Sb:SnSe.
The red and blue arrows indicate ${P_{+x}}$ and ${P_{-x}}$, respectively.
Only a few unit cells are shown for illustrating the ferroelectric barrier region
and electrode regions because of the page width limitations.
The lengths along $x$-axis in ${P_{+x}}$ and ${P_{-x}}$ states are both 114 \AA.
(b)
Calculated asymmetric potential energy profile as a function of
ferroelectric distortions in the 2D-FTJ In:SnSe/SnSe/Sb:SnSe.
The ferroelectric displacements $\lambda$ along [100] directions are
normalized so that
${\lambda = }$ +1 and -1 correspond to the ${P_{+x}}$
and ${P_{-x}}$ states, respectively.
The energy is summed up to all the atoms in the structure.
The energy of ${P_{-x}}$ state is set as the reference.
}
\end{figure}

In order to realize the functional 2D-FTJ $p$-SC/FE/$n$-SC, we turn to
study real materials.
SnSe is a 2D monochalcogenide with
robust in-plane ferroelectric polarization
above room temperature, as pointed out by Ref.~\citen{Fei2016PRL}.
The crystal structure of monolayer SnSe is explicitly provided
in the Supporting Information,
In the ${xy}$ plane, the ferroelectric polarization
along $x$ and $y$ axis are equivalent due to the symmetry~\cite{SM},
and we will only discuss the case along $x$-direction
throughout the study.
Using Berry phase
approach~\cite{PhysRevB.47.1651,RevModPhys.66.899},
we have revisited its polarization in the monolayer limit and found it is
$\sim$ 1.82$\times$10$^{-10}$ C/m, in agreement with
previous study~\cite{Fei2016PRL}.
In the Supporting Information, we also revisit the
double-well potential energy profile of monolayer SnSe and its lattice
dynamics properties.
As many other ferroelectric materials
like bulk BaTiO$_3$ and PbTiO$_3$~\cite{fang2015first},
monolayer SnSe also features a typical symmetric
potential profile~\cite{SM}, \ie,
the phase transition between its paraelectric and ferroelectric states
is continuous and
spontaneous below a critical temperature.
Utilizing different valences of elements,
In (Sb) is doped into monolayer SnSe to form $p$-type ($n$-type)
semiconductor. The experimental feasibility of such doped SnSe is
discussed in Supporting Information.
In our study,
the doped carrier concentration in In:SnSe or Sb:SnSe reaches up to 6.2$\times$10$^{20}$ cm$^{-3}$
because the size of unit cell In:SnSe or Sb:SnSe are composed of 4-unit cells SnSe where one Sn atom
is replaced by one In or Sb atom~\cite{SM}.
By examining the electronic structures of $p$-type In:SnSe and
$n$-type Sb:SnSe~\cite{SM},
the In and Sb dopants are found to introduce shallow levels.
Hence,
the doped holes and electrons are free to move in materials.
In addition, we find the density of states of In:SnSe is about two times of
that of Sb:SnSe, indicating the screening length of In:SnSe
is smaller than that of Sb:SnSe according to Thomas-Fermi theory
(details are provided in Supporting Information).
Therefore, In:SnSe and Sb:SnSe can act as semiconductor electrodes of
a 2D-FTJ.
We then set up a In:SnSe/SnSe/Sb:SnSe homostructure as a
implementation of $p$-SC/FE/$n$-SC 2D-FTJ.
In this In:SnSe/SnSe/Sb:SnSe homostructure,
the region of ferroelectric SnSe is a
stacking of 18 unit-cells along [100] direction ($\sim$ 8 nm),
and In:SnSe (Sb:SnSe) plays as the $p$-type ($n$-type)
semiconductor electrode.
The optimized atomic structures of In:SnSe/SnSe/Sb:SnSe homostructure
in the $P_{+x}$ and $P_{-x}$ states
are displayed in Figure~\ref{fig:Fig3}(a).
In addition to the robust ferroelectric displacements in the
barrier region, we find the
polar displacements remain in the In:SnSe and Sb:SnSe regions.
In contrast
to the conventional Metal-1/ferroelectric/Metal-2 FTJ~\cite{PhysRevLett.98.137201}
or all-oxide FTJ~\cite{velev2008magnetic} in which lattice mismatch
is inevitably generated by the electrodes and ferroelectrics due to
their unmatched lattice constants, our In:SnSe/SnSe/Sb:SnSe
homostructure completely eliminates the lattice constant mismatch
and will not introduce abrupt structure distortions at the
ferroelectric/electrode interfaces.
In addition,
the ferroelectric polarization of monolayer SnSe in this 2D-FTJ
is spontaneous, which avoids using sophisticated
chemical methods~\cite{YangQing-JACS-2017,WuMengHao-nanolett-2016}
to maintain stable
and switchable 2D ferroelectricity.

The stability and robustness of in-plane ferroelectricity in
a 2D ferroelectric barrier sandwiched between two doped electrodes
lies at the heart of 2D-FTJ device.
To verify the in-plane ferroelectricity of monolayer SnSe barrier
survives in the monolayer In:SnSe/SnSe/Sb:SnSe
homostructure, we compute the total energy of the homostructure
as a function of normalized polar displacements $\lambda$.
As shown in Figure~\ref{fig:Fig3}(b),
a double-well potential profile is observed, indicating the stability
of ferroelectricity in the monolayer SnSe. This can be ascribed to the
small depolarization field of ${6.5\times 10^6~\mathrm{V/m}}$ in SnSe barrier,
which can be well screened by the the left/right electrodes~(see Supporting Information for more details).
Different from a free-standing SnSe monolayer with a symmetric
potential profile~\cite{SM}, the monolayer In:SnSe/SnSe/Sb:SnSe homostructure displays
an asymmetric potential profile: 1) the energy minima at $\lambda$ = -1 (\ie,
$P_{-x}$ state)
and $\lambda$ = 1 ($P_{+x}$ state) are inequivalent
with energy difference of 161.25 meV/homostructure;
2) the energy maximum corresponding to a paraelectric phase
is approximately located
at $\lambda$ = 0.2, which slightly deviates from $\lambda$ = 0.
This asymmetry is a consequence of the symmetry breaking
introduced by the two different
dopants in the two electrodes.
The asymmetry can also be observed from the polar displacements profile in
Supporting Information.
These results indicate
the barriers at the two interfaces are also asymmetric, and will be responsible
for the TER effect~\cite{LLTao-2016APL-FTJ}.

\begin{figure}[!t]
\includegraphics[angle=0,width=0.99\textwidth]{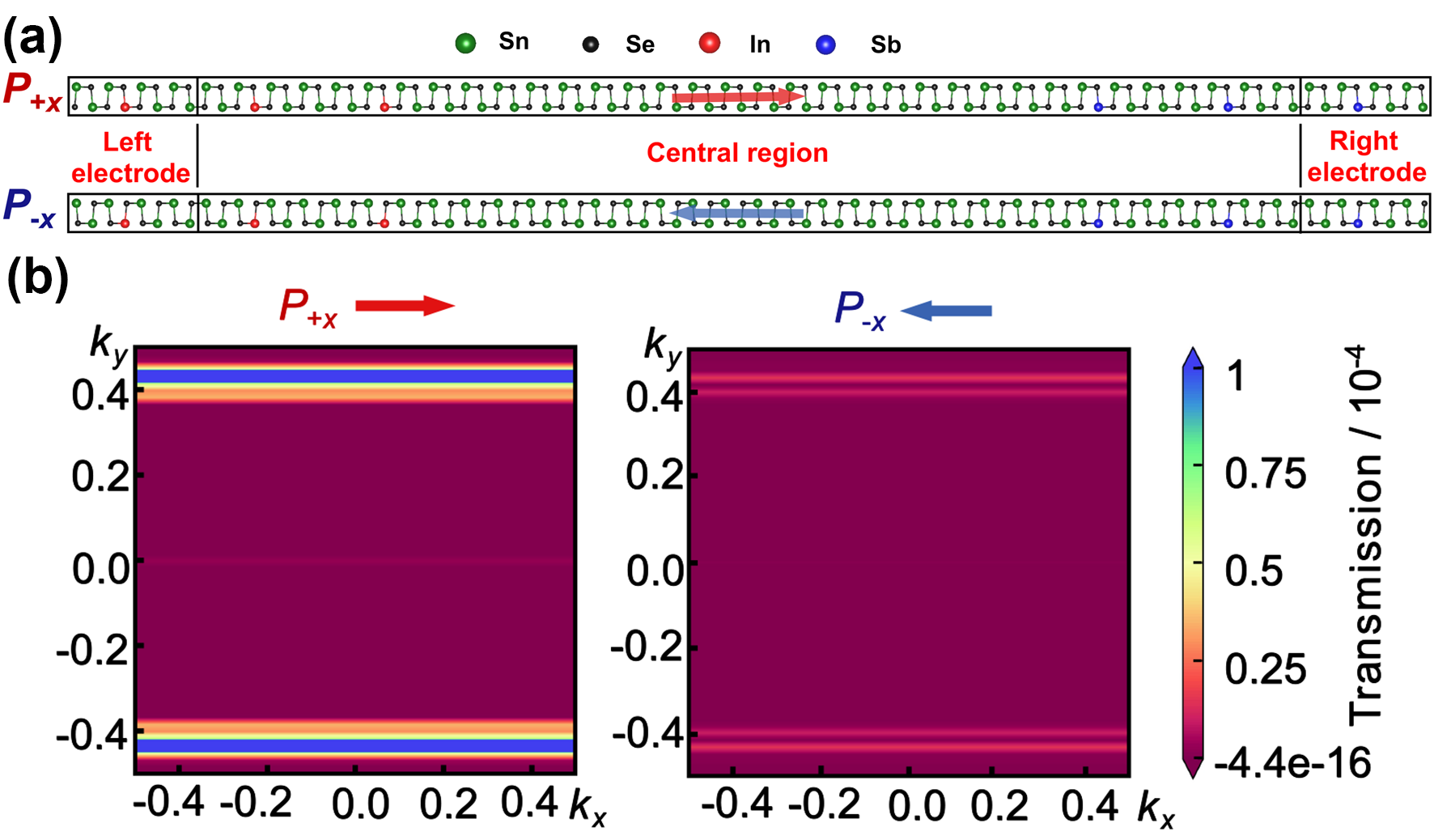}
\caption{\label{fig:Fig4}
(a) The device configurations in the DFT+NEGF calculations for
${P_{+x}}$ and ${P_{-x}}$ states.
(b) The $k_{\|}$-resolved transmissions in 2D Brillouin zone at
the Fermi energy through the 2D-FTJ In:SnSe/SnSe/Sb:SnSe
for ${P_{+x}}$ and ${P_{-x}}$ states.
Polarization directions are shown by arrows.
The $k_{\|}$-resolved transmissions for ${P_{+x}}$ and ${P_{-x}}$ states
use the same scale as given by the color bar.
}
\end{figure}

In order to evaluate the performance of 2D-FTJ In:SnSe/SnSe/Sb:SnSe,
density functional
theory plus non-equilibrium Green's function formalism
is used to
study the electrical conductance and TER effect.
The device configurations for
${P_{+x}}$ and ${P_{-x}}$ states are explicitly shown in Figure~\ref{fig:Fig4}(a).
The left/right extension layer (buffer layer)
is as wide as around 35 \AA, which is confirmed to be large enough to
screen the electrostatic potential~\cite{SM}.
In our calculations, the transmission coefficients and
reflection matrices are determined by matching
the wave functions of the scattering region with linear combinations of
propagating Bloch states in the electrodes.
Since the electronic states
at the ${E_{\rm F}}$ dominate the transport properties, the zero-bias electrical
conductance within the Landauer-B\"{u}ttiker formula~\cite{Landauer-1970} can be evaluated as:
\begin{equation}
        \label{Eq2}
    G = G_0 \sum_{k_\|} T(E_{\rm F}, k_\|) \,.
\end{equation}
where ${G_0 = 2e^2/h}$ is the conductance quantum, $e$ is the electron charge,
$h$ is the Planck's constant, and
${T(E_{\rm F}, k_{\|})}$ is the transmission coefficient at the Fermi energy
for a given Bloch wave vector ${k_{\|} = (k_x, k_y)}$ in the 2D
Brillouin zone.
By integrating the transmission probability
for states at the Fermi energy over the 2D Brillouin zone,
total conductance ($G$) can be calculated.
In the ${P_{+x}}$ state, $G_R$ = 1.003$\times$10$^{-9}$ S,
by contrast in the ${P_{-x}}$ state, $G_L$ = 6.435$\times$10$^{-11}$ S.
Following the conventional
definition in previous study~\cite{velev2008magnetic},
the TER ratio in our study is defined as:
\begin{equation}
        \label{Eq3}
        {\rm TER} = \frac{G_R - G_L}{G_L} \,.
\end{equation}
As a result, the reversal of ferroelectric polarization in the
2D-FTJ In:SnSe/SnSe/Sb:SnSe leads to a significantly enhanced TER
effect at zero bias, which is approximately about 1460$\%$.
This TER effect is about one order larger than those in conventional
all-oxide FTJs such as LaNiO$_3$/BaTiO$_3$/LaNiO$_3$~\cite{LLTao-2016APL-FTJ}
and SrRuO$_3$/BaTiO$_3$/SrRuO$_3$~\cite{velev2008magnetic} at zero bias.
Note that the migrations of electrons or holes near the semiconductor
surfaces are nearly ignored during the transport calculations in which only
electron cloud diffusion is taken into account, making the change of barrier
width underestimated by ferroelectric polarization reversal.
Hence, the actual TER effect in experiment should be even higher
than the theoretical value~\cite{wen2013ferroelectric}.
In addition, although the band gap of semiconductor
 is usually underestimated in DFT calculations,
we find the correction of band gap of ferroelectric barrier does not
affect the TER effect of the studied 2D-FTJ significantly~\cite{SM}, which is similar
to the conventional FTJs~\cite{TaoLL-JAP2016}.

In order to understand the large change in the conductance
ratio during the polarization reversal, the $k_{\|}$-resolved
transmissions
at ${E_{\rm F}}$ are shown in Figure~\ref{fig:Fig4}(b).
In the $P_{+x}$ state, the transmission coming
from the two blue stripe regions
(around $k_y$ = +0.42/-0.42) of the 2D Brillouin zone
are largest, indicating the feature of resonant tunneling.
We find the transmission eigenstates
around this region show much smaller decay rate than those
around $\Gamma$
point, which is responsible for the significant tranmission.
The discussions of transmission eigenstates are available in the Supporting
information.
Compared to the ${P_{+x}}$ state,
the transmission in the ${P_{-x}}$ state
is largely reduced, leading to lower conductance
than
the $P_{+x}$ state.
This explains the observed giant TER effect in
the monolayer In:SnSe/SnSe/Sb:SnSe homostructure.

\begin{figure}[!t]
\includegraphics[angle=0,width=0.9\textwidth]{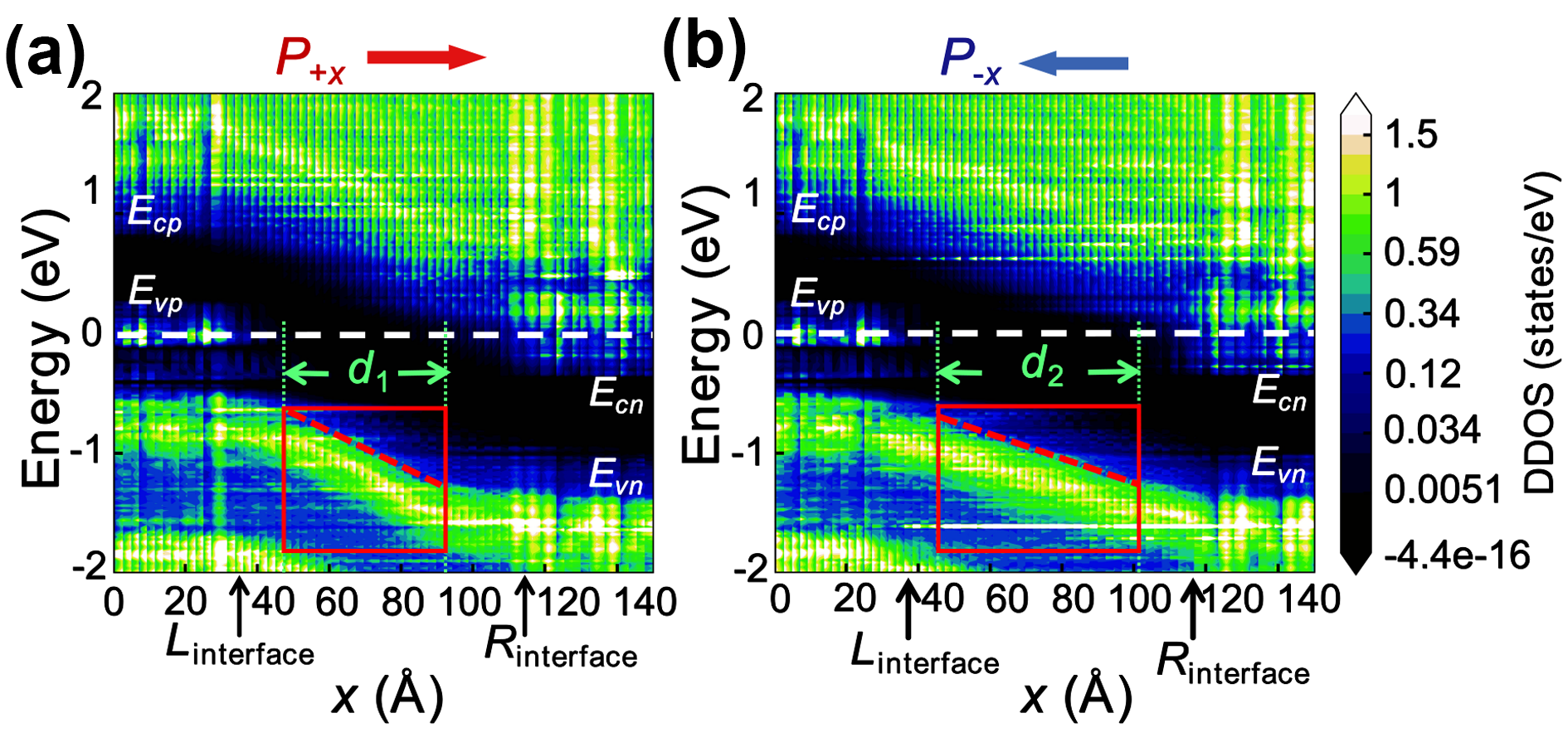}
\caption{\label{fig:Fig5}
The device density of states (DDOS) of the
2D-FTJ In:SnSe/SnSe/Sb:SnSe device projected onto its $x$-axis.
(a) ${P_{+x}}$ state. (b) ${P_{-x}}$ state.
The abscissa is the Cartesian coordinate of central region along the $x$-axis.
The color bar on the right indicates the DOS amplitude.
${E_{ci}}$ and ${E_{vi}}$ are the conduction band minimum and valence
band maximum of the electrodes, respectively.
The Fermi level is set to zero, which is shown by the white dashed line.
The interface of semiconductor/ferroelectric is initially set to be located around 35 \AA~and 114~\AA.
The black arrow indicates the interface between 2D ferroelectric SnSe and electrode In:SnSe (Sb:SnSe).
The red rectangle represents the region of effective ferroelectric tunneling barrier,
and the red dashed lines are used to guide
the evolution of the valence band maximum along $x$-axis.
The green arrow indicates the width of the tunneling barrier.
The black regions in the tunneling regions indicate the band gap of ferroelectric barrier.
}
\end{figure}

Finally, we compare the effective barriers for the two polarization
states in the 2D-FTJ In:SnSe/SnSe/Sb:SnSe.
As mentioned above, the combination of DFT-calculated density of states and Thomas-Fermi theory indicates
the screening length of left electrode In:SnSe
is smaller than the right electrode Sb:SnSe, this indicates the assumption that
${\delta_1 < \delta_2}$ in the model of Figure~\ref{fig:Fig2}
is indeed true in our studied case.
We can then get ${\mathit{\Phi}_R}$ $<$ ${\mathit{\Phi}_L}$ through Equations~\ref{Ineq:screen} and~\ref{Eq:overallbarrier},
\ie, the barrier height for ${P_{+x}}$ state
is smaller than that for ${P_{-x}}$ state.
In order to understand the change of barrier width,
we study the electronic structure across the 2D-FTJ In:SnSe/SnSe/Sb:SnSe
device by carrying out the analysis of real-space device DOS (DDOS) projected onto
the device $x$-axis.
The corresponding
results for the two polarization states are displayed in Figure~\ref{fig:Fig5}.
We find that the band diagrams calculated by DFT+NEGF approach are qualitatively consistent with
those obtained in our model in Figure~\ref{fig:Fig2}.
In particular,
barrier width in the $P_{-x}$ state is indeed increased compared to
the $P_{+x}$ state owing to the
interfacial metallization of the edge regions in ferroelectric barrier controlled by the
ferroelectric switching.
In Figure~\ref{fig:Fig5}, we have used red solid rectangles
to highlight the
effective tunneling regions,
in which the evolution of the valence band maximum along $x$-axis
is guided by the red dashed line.
The built-in electric field caused by the
work function step
can be clearly observed through the tilting of bands in
SnSe~\cite{PhysRevLett.98.207601}.
When the ferroelectric polarization is pointing to
the right (\ie, ${P_{+x}}$ state),
the depolarizing
field is parallel to the built-in electric field,
and hence the band edges in SnSe is
tilted.
On the other hand, the bands of SnSe become slightly
flat since the depolarization field is antiparallel to the
built-in field
in the ${P_{-x}}$ state.
More details about the reversible
metallization of the monolayer SnSe barrier
can be found in the
layer-resolved density of states provided in
Supporting Information.
Therefore, in addition to the raised
barrier height as polarization is flipped from ${P_{+x}}$ to ${P_{-x}}$
state, the barrier width is also increased. This makes the electron tunneling
much easier in ${P_{+x}}$ state than in ${P_{-x}}$ state,
which is account for the observed giant TER effect.

\section*{CONCLUSIONS}
In summary,
we have proposed a method to design two-dimensional ferroelectric tunnel junction
based on planar $p$-type semiconductor/ferroelectric/$n$-type
semiconductor homostructure via the doping engineering in a two-dimensional ferroelectric semiconductor with
in-plane electric polarization.
Combining density functional theory calculations with non-equilibrium Green's function formalism,
a giant TER effect of 1460$\%$~is observed in our newly designed 2D-FTJ In:SnSe/SnSe/Sb:SnSe homostructure, which
is comparable to that of conventional all-oxide FTJs.
The tunable tunneling barrier width that is generally
absent in conventional FTJs,
as well as the tunneling barrier height, is responsible for
the enhanced TER effect.
The dynamical modulation of barrier width stems from the
depletion/accumulation of majority carriers near the
semiconductor surface in response to the reversal of ferroelectricity.
SnSe is thought to be the most flexible
in known 2D atomic materials~\cite{zhang2016tinselenidene}, hence monolayer In:SnSe/SnSe/Sb:SnSe homostructures
can be promising memory blocks in the wearable devices and artificial synapses,
which has larger advantages
over the conventional FTJs.
The proposed strategy in our study is applicable to design novel 2D-FTJs
using other 2D ferroelectric materials.
We hope this work
will stimulate the experimental endeavors of fabricating
2D-FTJs with giant TER effect to accelerate their commercial applications
into ultra-low-power, high-speed,
and non-volatile nanoscale memory devices.

\section*{Supporting Information}
The Supporting Information can be found after the main text.
	It is also available free of charge on the
ACS Publications website at DOI: 10.1021/acsaelm.9b00146.

    \begin{itemize}
	\item The overall potential energy profile if ${\delta_1 > \delta_2}$; in-plane
       ferroelectricity in monolayer SnSe; carrier concentration in doped SnSe;
        phonon dispersions; dopants with shallow levels;
        screening lengths in electrodes; experimental feasibility;
        depolarization field; polar displacements profile;
        layer-resolved density of states; transmission eigenstate;
        width of extension layer; Hubbard $U$ effect
\end{itemize}

\section*{acknowledgement}
    This work was supported by the National Key R$\&$D Program of China (2017YFA0303403),
    the National Natural Science Foundation of China (Grant No. 11774092, 51572085)
    and Shanghai Science and Technology Innovation Action Plan (No. 17JC1402500).
    Computations were performed at the ECNU computing center.


\textbf{Competing interests:}
The authors declare no competing financial or non-financial interests.

\bibliography{2D_FTJ}

\end{document}


\title{Supporting Information for ``Two-dimensional ferroelectric tunnel junction:
  the case of monolayer In:SnSe/SnSe/Sb:SnSe homostructure''
  }

\author{Xin-Wei Shen$^{1}$, Yue-Wen Fang$^{1,2*}$, Bo-Bo Tian$^{1}$, Chun-Gang Duan$^{1,3}$}
\email{fyuewen@gmail.com (Y.-W.F); cgduan@clpm.ecnu.edu.cn (C.-G.D)}
\affiliation{
$^{1}$State Key Laboratory of Precision Spectroscopy and Key Laboratory of Polar Materials and Devices, Ministry of Education, Department of Optoelectronics, East China Normal University, Shanghai, 200241, China \\
$^{2}$Department of Materials Science and Engineering, Kyoto University, Kyoto 606-8501, Japan \\
$^{3}$Collaborative Innovation Center of Extreme Optics, Shanxi University, Taiyuan, Shanxi 030006, China \\
}

\maketitle

  \tableofcontents
\makeatletter
\let\toc@pre\relax
\let\toc@post\relax
\makeatother
    \clearpage
  \newpage

  \section{The overall potential energy profile if ${\delta_1 > \delta_2}$}
\begin{figure}[!htp]
\includegraphics[angle=0,width=0.7\textwidth]{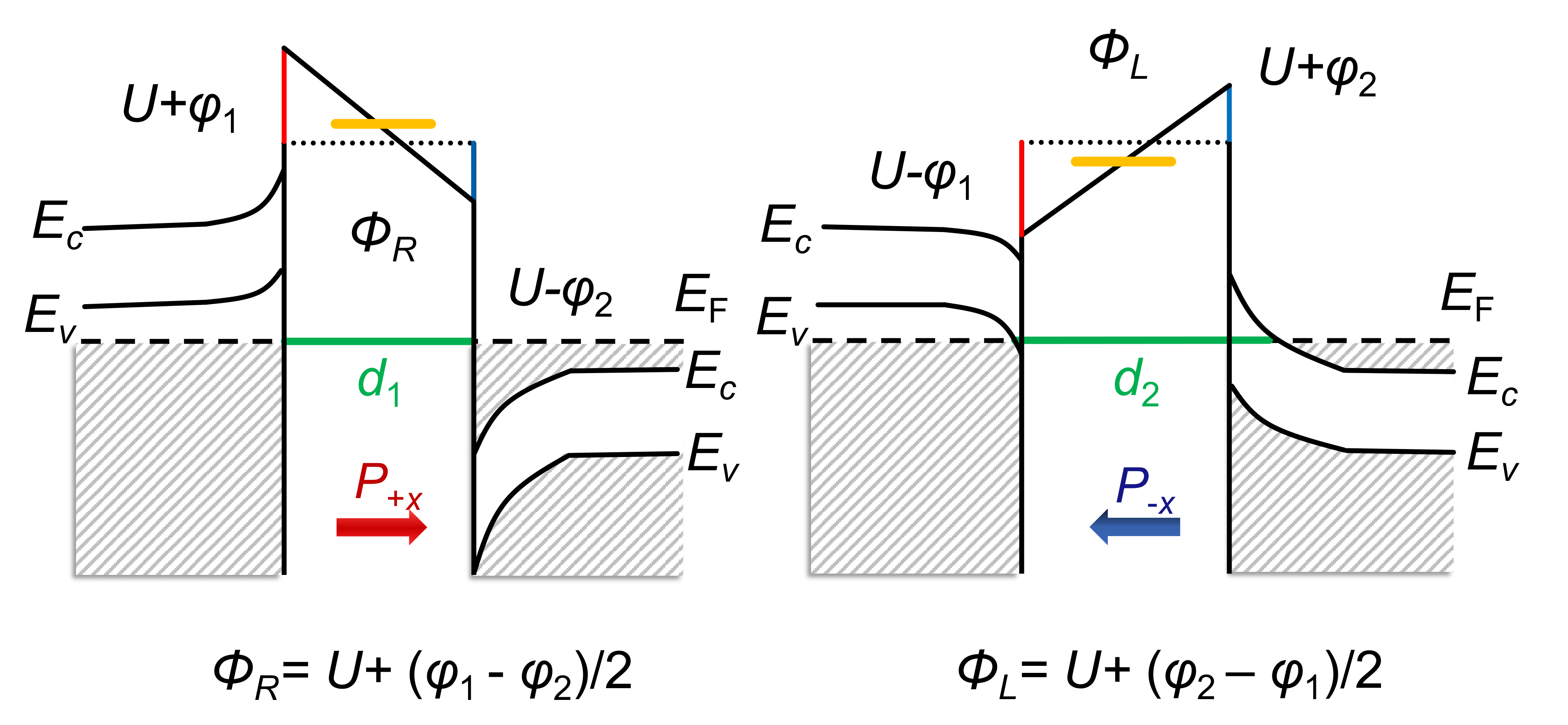}
\caption{\label{fig:model-supp}
Switching mechanisms of the 2D-FTJ $p$-SC/FE/$n$-SC.
The overall potential energy profiles with corresponding band diagrams if we assume ${\delta_1 > \delta_2}$.
Here, ${\delta_1}$ and ${\delta_2}$ are
the screening lengths of left and right electrodes,
respectively.
The barrier width in either ${P_{+x}}$ or ${P_{-x}}$ state is given by the
green line.
The average potential barrier height is indicated by the short orange line.
The Fermi level, as shown by black dashed line, is set as the reference of the barrier height.
}
\end{figure}

Figure~\ref{fig:model-supp} shows the overall potential energy profiles
if we assume ${\delta_1 > \delta_2}$ (${\delta_1}$ and ${\delta_2}$ are
the screening lengths of left and right electrodes,
respectively).

\section{The in-plane ferroelectricity in monolayer SnSe}
\begin{figure}[!htp]
\includegraphics[angle=0,width=0.65\textwidth]{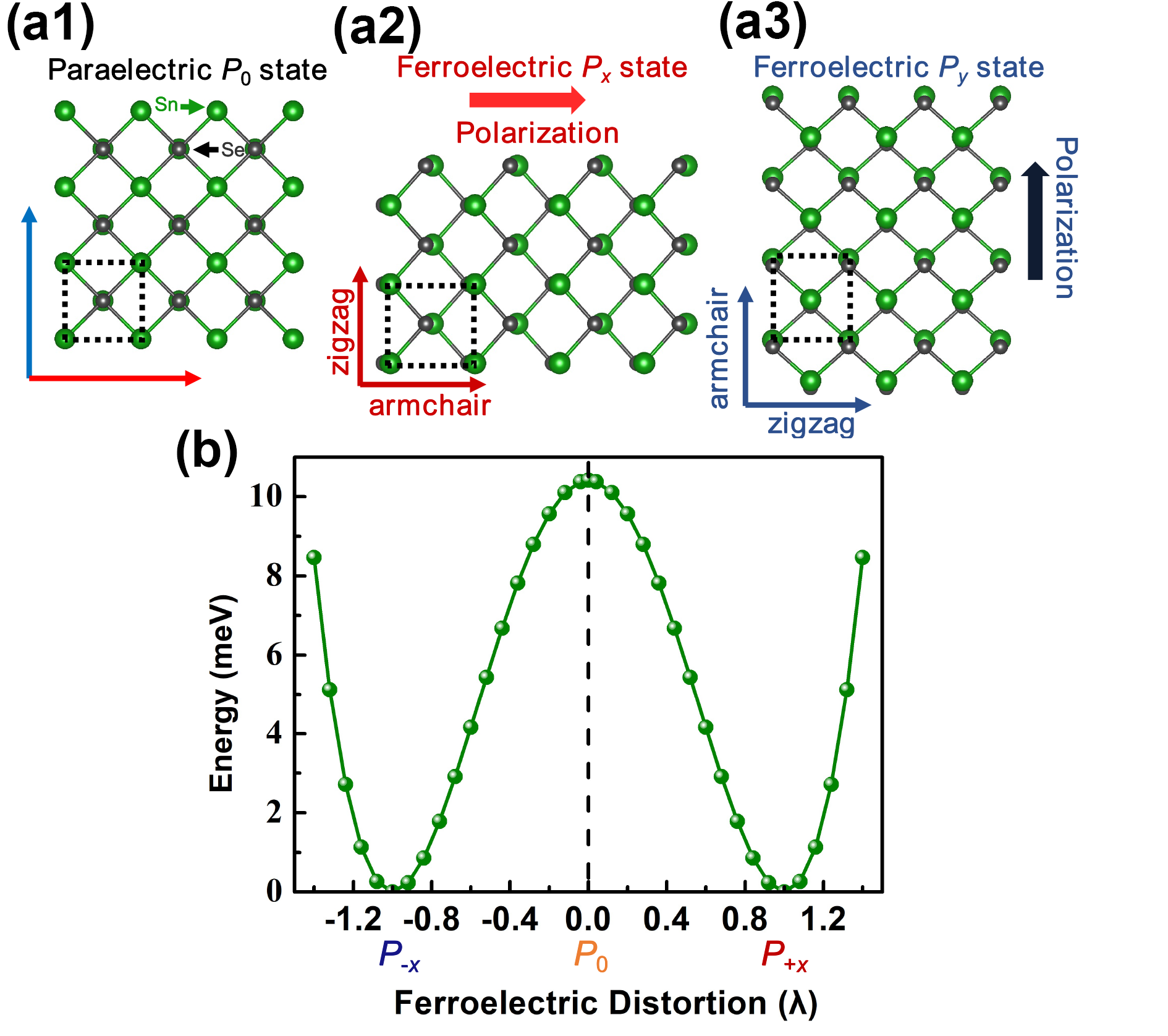}
\caption{\label{fig:s1}
The top views of lattice structure for the monolayer SnSe and the double well potential.
(a1) The paraelectric state.
(a2) The ferroelectric state along $x$-direction.
(a3) The ferroelectric state along $y$-direction.
The black dashed rectangle represents the primitive cell.
(b) Calculated double-well potential energy profile
of monolayer SnSe.
The amplitude $\lambda$ of ferroelectric displacements along [100] directions
is normalized, where -1 (+1) corresponds to the
${P_{-x}}$ (${P_{+x}}$) state,
and 0 corresponds to the paraelectric state.
}
\end{figure}

The top view of paraelectric SnSe monolayer (${P_0}$ state) is
illustrated in Figure~\ref{fig:s1}(a1),
showing a high-symmetry lattice structure with
the space group of ${Cmcm}$.
In this centrosymmetric ${Cmcm}$ structure, the symmetry determines the
atomic arrangements along $x$-axis and $y$-axis are exactly same.
Hence, the corresponding noncentrosymmetric ferroelectric phase with polarization
along $x$-axis (referred to as $P_x$ state) is exactly equivalent to the one with polarization along $y$-axis
(referred to as $P_y$ state),
as explicitly shown in Figure~\ref{fig:s1}(a2) and (a3).
We note that the armchair direction is always the direction of ferroelectric polarization.
The space group of $P_x$ or $P_y$ state is ${Pnma}$.
In the following sections and the main text, we only discuss $P_x$ state (\ie, along [100] direction) as an example.

Figure~\ref{fig:s1}(b) shows the
calculated double-well potential profile of the pure monolayer SnSe.
In this pure phase without doping,
the energy potential profile shows a symmetric
feature,
\ie, $P_{+x}$ and ${P_{-x}}$ are equivalent.
The energy difference between the paraelectric and ferroelectric
states is estimated to be 10.42 meV.
The smooth energy curve connected paraelectric phase with the ferroelectric phase
indicates the existence of a spontaneous symmetry breaking, in other words,
the ferroelectric polarization is spontaneous.

  \section{The carrier concentration in doped SnSe}
  \label{carrier-concentration}
In our calculations, the optimized lattice constants of the primitive cell for pristine SnSe
are a = 4.38 \AA, b = 4.29 \AA, and c = 21.46 \AA, respectively.
The unit cell size of electrode (\ie, doped SnSe) is a stacking of 4 unit-cells SnSe along $x$-axis in
which one Sn atom is substituted by one In or Sb atom.
Therefore, the size of unit cell doped SnSe is about 1612.95 \AA$^3$,
and the doping
concentration can be calculated as 1/1612.95 = 6.20$\times$10$^{20}$ cm$^{-3}$.

  \section{Phonon dispersions}
\begin{figure}[!htp]
      \includegraphics[angle=0,width=0.8\textwidth]{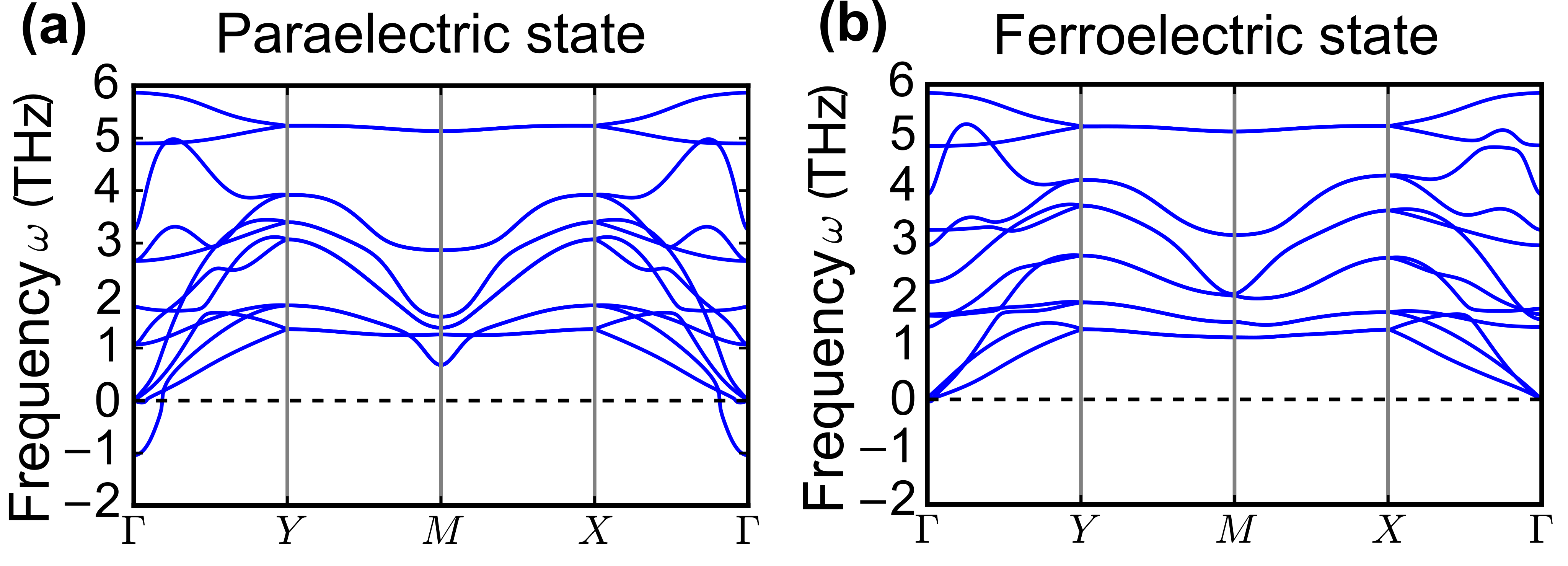}
      \caption{\label{fig:s3}
      Phonon dispersion curves along the high-symmetry $k$ path
      for the monolayer SnSe.
      (a) The paraelectric state with unstable mode.
      (b) The ferroelectric ground-state.
      }
  \end{figure}

  Figure~\ref{fig:s3}(a) shows the phonon dispersion of paraelectric
  state of monolayer SnSe. Significant imaginary modes
  appearing at $\Gamma$ point are observed, indicating the
  large ferroelectric instability.
  The emerging unstable soft mode leads to the energy-lowering
  ferroelectric distortions of SnSe and the stabilization
  of ferroelectric state.
  The corresponding phonon dispersion of ferroelectric state
  is shown in Figure~\ref{fig:s3}(b).
  By comparing our phonon calculations with that reported by
  Fei \etal~\cite{Fei2016PRL}, we find
  the phonon dispersions share similarities but also
  have minor differences.
  Specifically, we observe some differences of phonon
  modes around the high symmetry points of $\Gamma$ and $M$.
  The differences should mainly come from two reasons:
  1) different lattice constants;
  2) we note that the non-analytic contributions which
  are important for ferroelectric materials are not taken
  into consideration in Fei \etal's study~\cite{Fei2016PRL}.
  This leads to the slight difference in
  LO-TO splittings between their study and ours.

\section{Dopants with shallow levels}
  \label{Dopants-levels}
\begin{figure}[!htp]
\includegraphics[angle=0,width=0.8\textwidth]{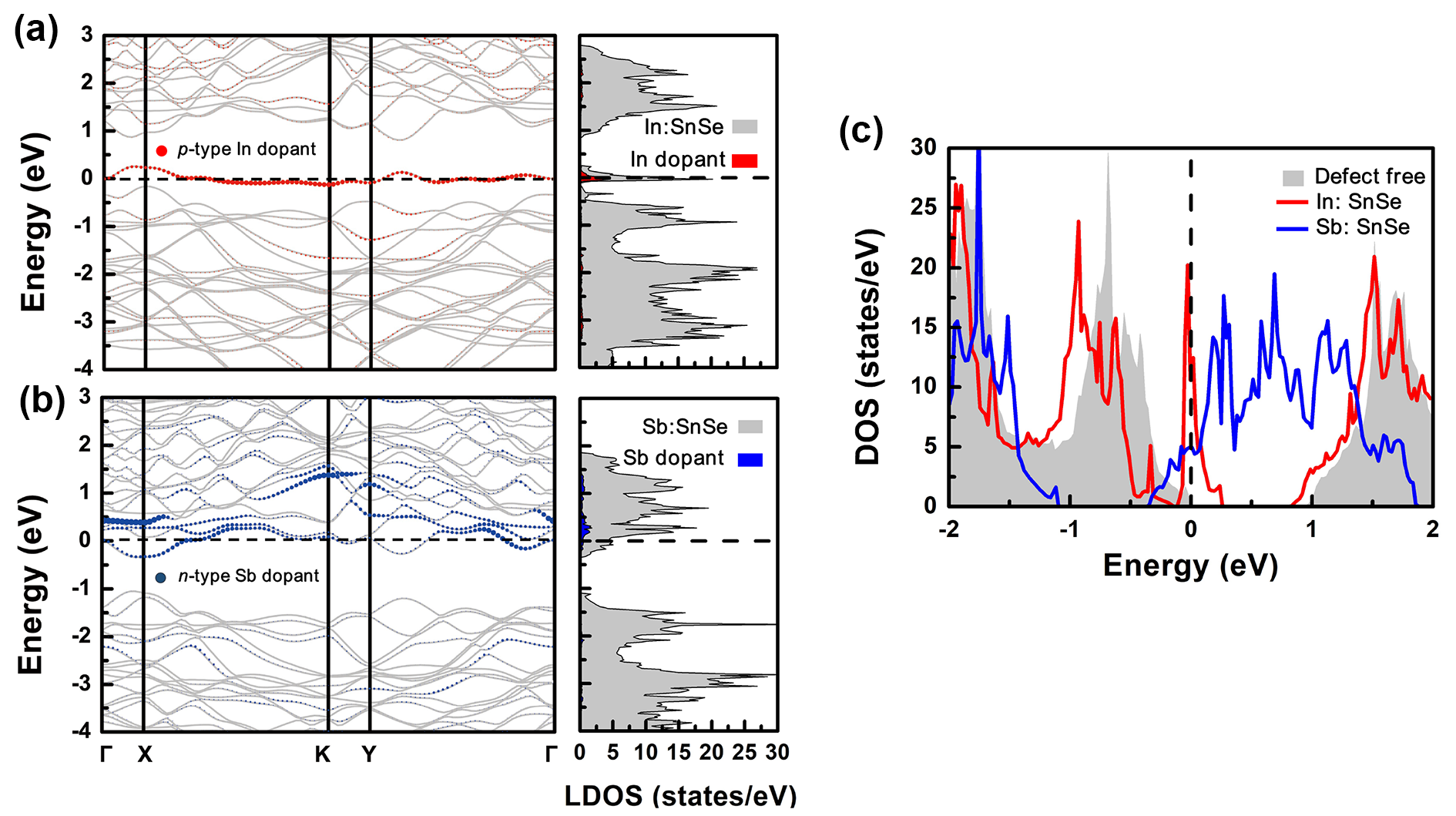}
\caption{\label{fig:s2}
Dopant-weighted band structure and the corresponding density of states (DOS) for
(a) $p$-type In:SnSe and
(b) $n$-type Sb:SnSe.
The circles indicate the weighted contributions of dopants to the bands.
(c) A comparison of the total DOS for In:SnSe, Sb:SnSe, and SnSe.
The Fermi level is set at 0 eV.
}
\end{figure}

Generally in the case of shallow donor or acceptor,
the electronic states of defect introduced by the
dopants will not be more than a few $k_B$T below the
conduction band minimum (CBM) or above the valence
band maximum (VBM) edge, respectively~\cite{RevModPhys.86.253}.
In order to examine the defect-induced levels in the doped SnSe,
we have calculated the electronic structures
of monolayer In:SnSe and Sb:SnSe, the results are shown in
Figure~\ref{fig:s2}(a) and (b).
For the $p$-type In:SnSe where
In atoms act as the acceptors, the band structure and
the corresponding density of states (DOS)
show a heavily doped condition,
in which VBM lies beyond the Fermi level.
  We show the weighted contributions of In dopants to
  electronic states using red solid circles,
  and find that the
components of the defect-induced states appear predominantly
near the VBM, therefore showing the key feature of a shallow
acceptor level.
In analogy to In:SnSe,
we find Sb:SnSe also provides a shallow donor level.
As a comparison, we show the total DOS for
In:SnSe, Sb:SnSe, and SnSe in Figure~\ref{fig:s2}(c).


\section{Screening lengths of In:SnSe and Sb:SnSe electrodes}
  \label{screening_length}

We estimate that the screening length of left electrode In:SnSe
is smaller than the right electrode Sb:SnSe
according to the Thomas-Fermi theory.

Specifically in Thomas-Fermi theory, the screening length of a metallic material
is determined by the materials' dielectric permittivity and electronic
states:
\begin{equation}
        \label{Eq:screen-length}
\delta = \frac{1}{e} \sqrt{\frac{\varepsilon}{\rho}}
\end{equation}
where $\varepsilon$ is the dielectric permittivity and $\rho$
is the density of states at the Fermi level ${E_{\rm F}}$.
The dielectric permittivities of In:SnSe and Sb:SnSe are equivalent
to that of SnSe at saturation polarization~\cite{PhysRevLett.109.247601}.
Hence, the screening lengths of the two electrodes
only depend on their densities of states at the ${E_{\rm F}}$ that
can be easily controlled by doping.
In our study, we only study the same $n$-type and $p$-type doping concentration, \ie,
6.2$\times$10$^{20}$ cm$^{-3}$ (see Section~\ref{carrier-concentration}).
As shown in Figure~\ref{fig:s2},
it can be found that the
density of states of In:SnSe at the Fermi level is about two times of
that of Sb:SnSe~\cite{SM}, indicating the screening length of In:SnSe
is smaller than that of Sb:SnSe.

\section{Experimental feasibility for achieving doped SnSe}
 According to Ref.~\cite{RevModPhys.86.253}, 
the formation energy of a defect in a charge state $q$
writes:
\begin{eqnarray}
\label{eq:formation}
\Delta H_f(\alpha, q) = E_T({\rm defect}) - E_T({\rm perfect}) + \sum_\alpha n_\alpha(\delta \mu_\alpha + \mu_\alpha^{\rm solid}) + q(E_v + E_{\rm F})\,,
\end{eqnarray}
where ${E_T({\rm defect})}$ and ${E_T({\rm perfect})}$
represent the total energies of the supercell with and without a defect, respectively.
The third term of Equation~\ref{eq:formation}
corresponds to the energy change due to exchange of atoms with the chemical reservoirs,
where $\alpha$ determines which atom is added or removed for the defect. If an atom is added,
${n_\alpha = -1}$, while ${n_\alpha = 1}$ if if an atom is replaced.
${\delta \mu_\alpha}$ is the elemental chemical potential of atom
$\alpha$, referenced to the total ground-state solid
${\mu_\alpha^{\rm solid}}$ of the pure elemental phase.
The last term of Equation~\ref{eq:formation}
is the energy change due to exchange of electrons and holes with the carrier reservoirs,
in which ${E_v}$ is set to the energy at the valence-band maximum
of the defect free system, and ${E_{\rm F}}$ is
the Fermi energy of the system measured from the valence-band maximum.

For a neutral ($q$ = 0) cation defect in the $p$-type doped left electrode,
the formation energy with $n_{\rm In}$ = $n_{\rm Sn}$  = 1
(\ie, the case of one Sn atom replaced by In atom)
can be calculated as following:
\begin{eqnarray}
\label{eq:formation-L}
\Delta H_f^L = E_T({\rm In:SnSe}) - E_T({\rm SnSe}) - \mu_{\rm In}^{\rm solid} - \delta\mu_{\rm In} + \mu_{\rm Sn}^{\rm solid} + \delta\mu_{\rm Sn}\,.
\end{eqnarray}
As a result, we can calculate the formation energy ${H_f^L}$ = 0.90 eV.
Similarly, for a neutral ($q$ = 0) anion effect in the $n$-type doped
right electrode, the formation energy is given by:
\begin{eqnarray}
\label{eq:formation-R}
\Delta H_f^R = E_T({\rm Sb:SnSe}) - E_T({\rm SnSe}) - \mu_{\rm Sb}^{\rm solid} - \delta\mu_{\rm Sb} + \mu_{\rm Sn}^{\rm solid} + \delta\mu_{\rm Sn}\,.
\end{eqnarray}
Thus, we get ${H_f^R}$ = 1.48 eV.

In order to evaluate the feasibility of the doping,
we make a comparison of the formation energy between the doped electrodes adopted here and previous works.
For example, the native defects in single-layer MoS$_2$ have been investigated in Ref.~\cite{PhysRevB.89.205417}.
We can find that in the Mo-rich limit condition, the S vacancy ($V_{\rm S}$) is found to be the most stable,
of which the formation energy is only about 1.5 eV for ${V_{\rm S}^0}$  and about 1.2 eV for ${V_{\rm S}^{1-}}$.
For the S-rich single-layer MoS$_2$, the S interstitial (S$_i$) is found the most stable with the formation
energy only about 1.0 eV.
Moreover, according to the stability of point defects in monolayer rhenium disulfide discussed in
Ref.~\cite{PhysRevB.89.155433}, we can find that the stable S-vacancy formation energies range from 1.16 to 2.39 eV
at the Re or S-rich condition.
On account of these results, we can see that such formation energies for both left and right electrodes in our works
confirm the feasibility of doping, which is of great possibility to achieve In or Sb doped SnSe
  in experiment.


%

  \section{Depolarization field in In:SnSe/SnSe/Sb:SnSe}
\begin{figure}[!htp]
	\includegraphics[angle=0,width=0.7\textwidth]{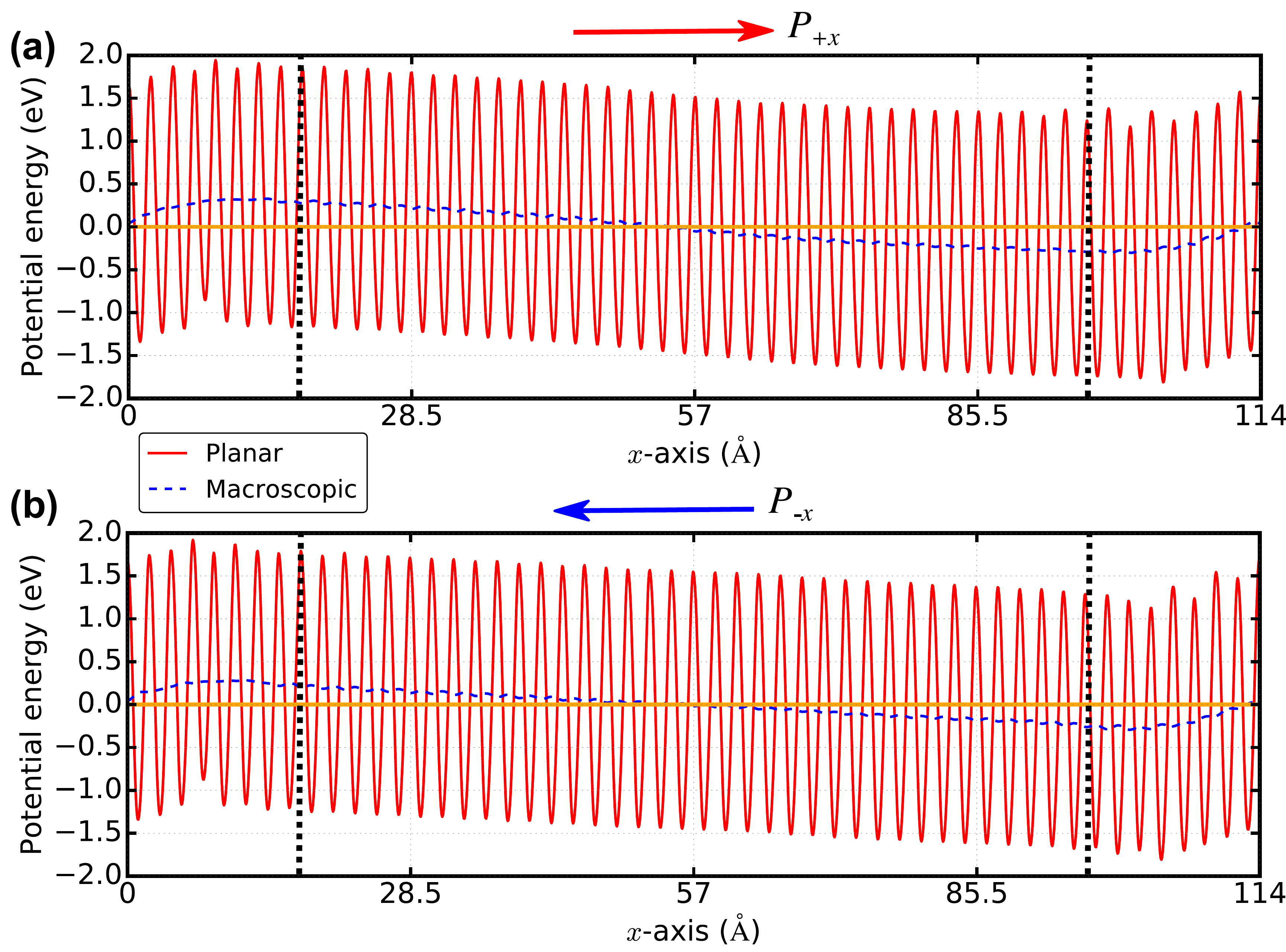}
\caption{\label{fig:LOCPOT}
The DFT-calculated electrostatic potential profiles of (a) $P_{+x}$ and (b) $P_{-x}$ states.
The solid red curve and dashed blue curve refer to planar average
and macroscopic average potentials, respectively.
The vertical dashed black lines denote the interfaces.
}
\end{figure}

We performed
analysis of the electrostatic potential of the supercell that
has been used in our double well potential calculations in the main text.
The DFT-calculated electrostatic potential energy profiles for the $P_{+x}$ and
$P_{-x}$ states are both shown in Figure~\ref{fig:LOCPOT}.
  The asymmetry in potential profile can be observed, which is in agreement with
  the asymmetric features in the double-well potential profiles in our main text.
The electrostatic potential energy presented in Figure~\ref{fig:LOCPOT}
implies the contribution from the combination of
(1) built-in electric field $E_{\rm bi}$ owing to the
asymmetric interface, and
(2) the depolarization field $E_d$ owning to the incomplete screening~\cite{Luo-JAP2012}.
Different from $E_{\rm bi}$ which is invariant during the polarization reversal,
$E_{\rm d}$ changes its direction because it is always opposite to
the direction of ferroelectric polarization.
By comparing the macroscopic average potential energy profiles,
we find the slope in the SnSe regions in the $P_{+x}$
state (\ie, panel (a) of Figure~\ref{fig:LOCPOT})
is larger than that in the $P_{-x}$ state, this implies
$E_{\rm bi}$ and $E_d$ have the same direction in the $P_{+x}$
state, while they have opposite directions in the  $P_{-x}$ state and
result in smaller electric field.
Specifically, we get
\begin{eqnarray}
    \begin{cases}
\label{Eq:E_bi-E_d}
 E_{\rm bi} + E_{\rm d} = 7.2 \times 10^7~\mathrm{V/m}\\
 E_{\rm bi} - E_{\rm d} = 5.9 \times 10^7~\mathrm{V/m}\,.
   \end{cases}
\end{eqnarray}
Therefore,  $E_{\rm bi}$ and ${E_{\rm d}}$
are solved to be ${6.6\times 10^7~\mathrm{V/m}}$ and
${6.5\times 10^6~\mathrm{V/m}}$, respectively.
The depolarization field in general ferroelectric films
is around ${10^8 \sim 10^9~\mathrm{V/m}}$~\cite{lang2007frontiers,de1996ferroelectric},
hence the very small depolarization field (${6.5\times 10^6~\mathrm{V/m}}$)
in our study implies
that the bound charge at the edges of ferroelectric
regions can be well screened by the left/right electrodes.
This indicates the ferroelectricity in the monolayer SnSe
between the electrodes is very stable, which is in agreement with
the double-well potential profile in Figure 3(b) in the main text.


\section{Polar displacements profile}
\begin{figure}[!htp]
	\includegraphics[angle=0,width=0.70\textwidth]{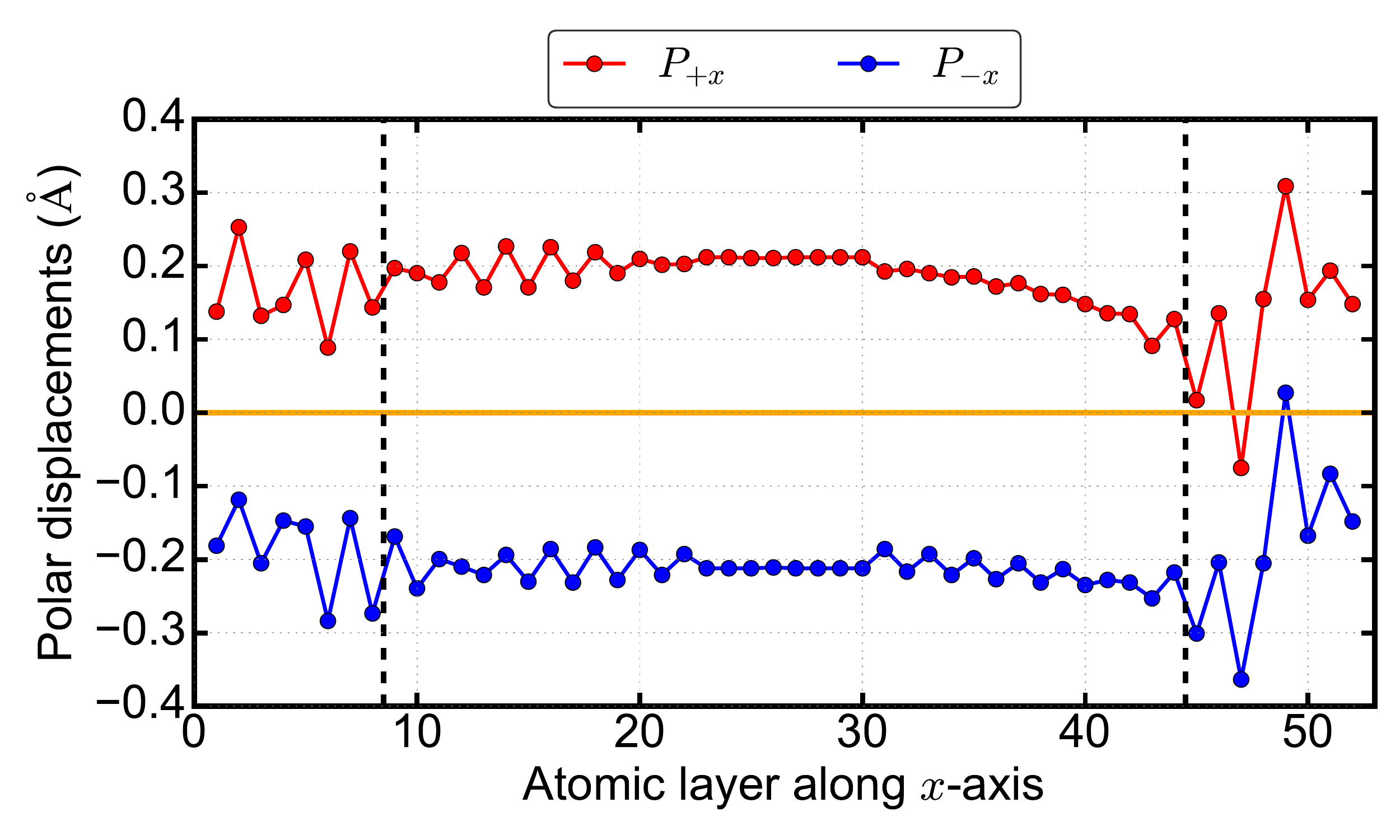}
\caption{\label{fig:disp}
The polar displacements profiles for ${P_{+x}}$
and ${P_{-x}}$ states of
the In:SnSe/SnSe/Sb:SnSe junction.
The measured polar displacements here refer to
the displacements of cations with respect to the anions.
The interfaces between the electrodes and barrier region are
indicated by vertical dashed lines.
}
\end{figure}

 Figure~\ref{fig:disp} shows the DFT-calculated polar displacements profile for both ${P_{+x}}$
and ${P_{-x}}$ states.

\clearpage
\newpage

\section{Layer-resolved density of states of In:SnSe/SnSe/Sb:SnSe}
\begin{figure}[!htp]
\includegraphics[angle=0,width=0.55\textwidth]{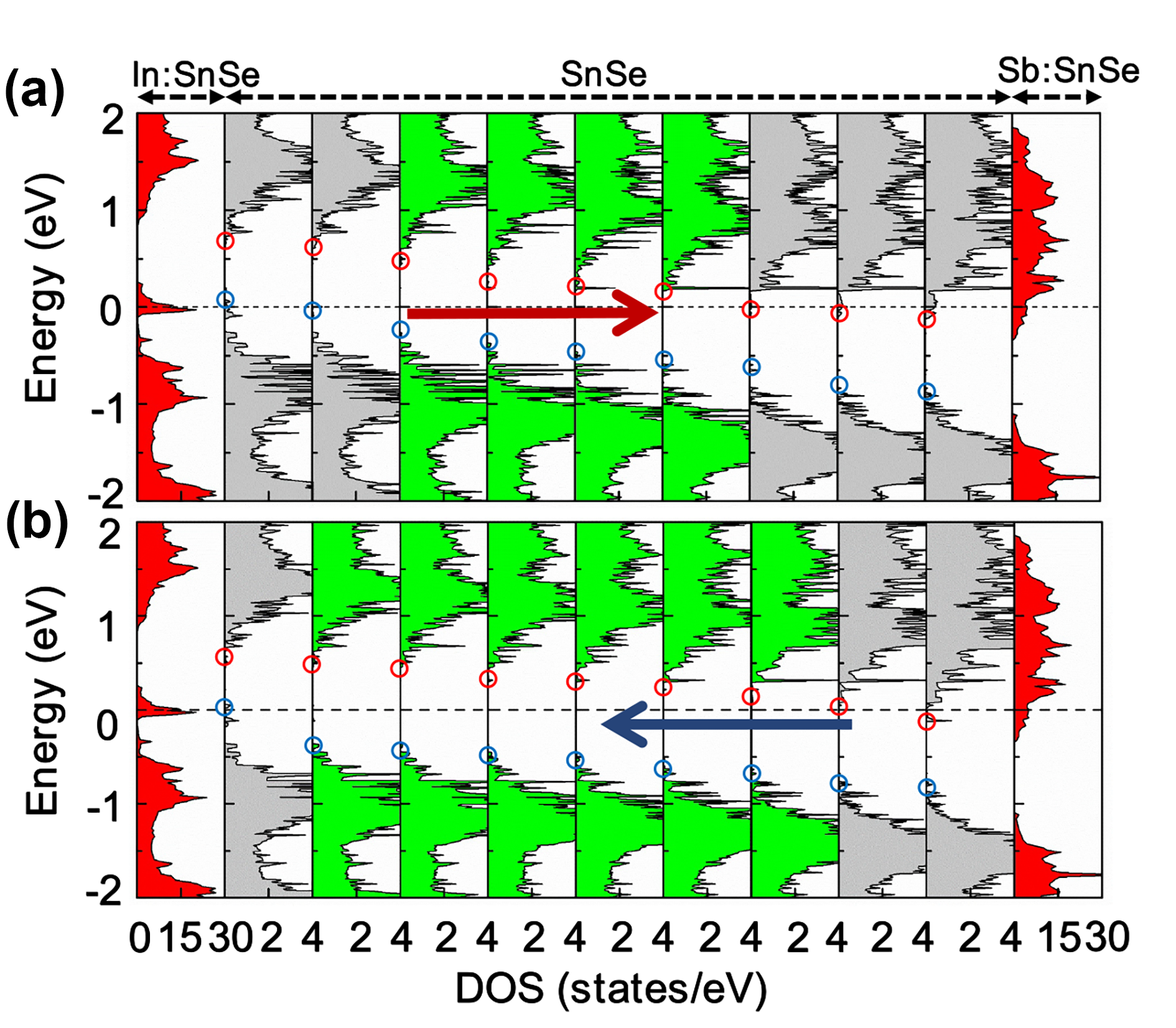}
\caption{\label{fig:s4}
Layer-resolved density of states for
(a) $P_{+x}$ and (b) $P_{-x}$ state.
The red and blue circles indicate the position of the CBM
and VBM, respectively.
The Fermi level is set to zero.
The red shaded areas denote the electrode regions,
the grey shaded areas represent the regions in SnSe barrier that become metallic,
the green shaded ones correspond to
the effective tunneling barrier.
}
\end{figure}

Figure~\ref{fig:s4} shows the DFT-calculated layer-resolved density of states of
In:SnSe/SnSe/Sb:SnSe junction, here layer in SnSe barriers
refers to two unit cells along $x$-axis.
In the In:SnSe (or Sb:SnSe) electrode, the densities of states
contributed by all the unit cells are summed
to make the demonstration clear.
As shown in Figure~\ref{fig:s4},
more SnSe unit cells in $P_{+x}$ state become
metallic than $P_{-x}$ state.
Consequently, the effective tunneling barrier width
in $P_{-x}$ state is wider than $P_{+x}$ state.

  \section{The analysis of transmission eigenstate}
  We note
  that the transmission is mainly contributed by the region around
  $k_y$ = +0.42/-0.42
  instead of around $\Gamma$ point. In order to understand why the resonant tunneling
  occurs in these special $k$ points instead of $\Gamma$ point,
  we further study the
  transmission eigenstates at the Fermi level for
  $k$ = (0, -0.42) and $k$ = (0, 0) (\ie, $\Gamma$ point).
  Since the transmission eigenstates
  correspond to the scattering state from the left electrode to the right electrode,
  we can find the electronic states that contribute to the electron transport.

\begin{figure}[!htp]
      \includegraphics[angle=0,width=0.75\textwidth]{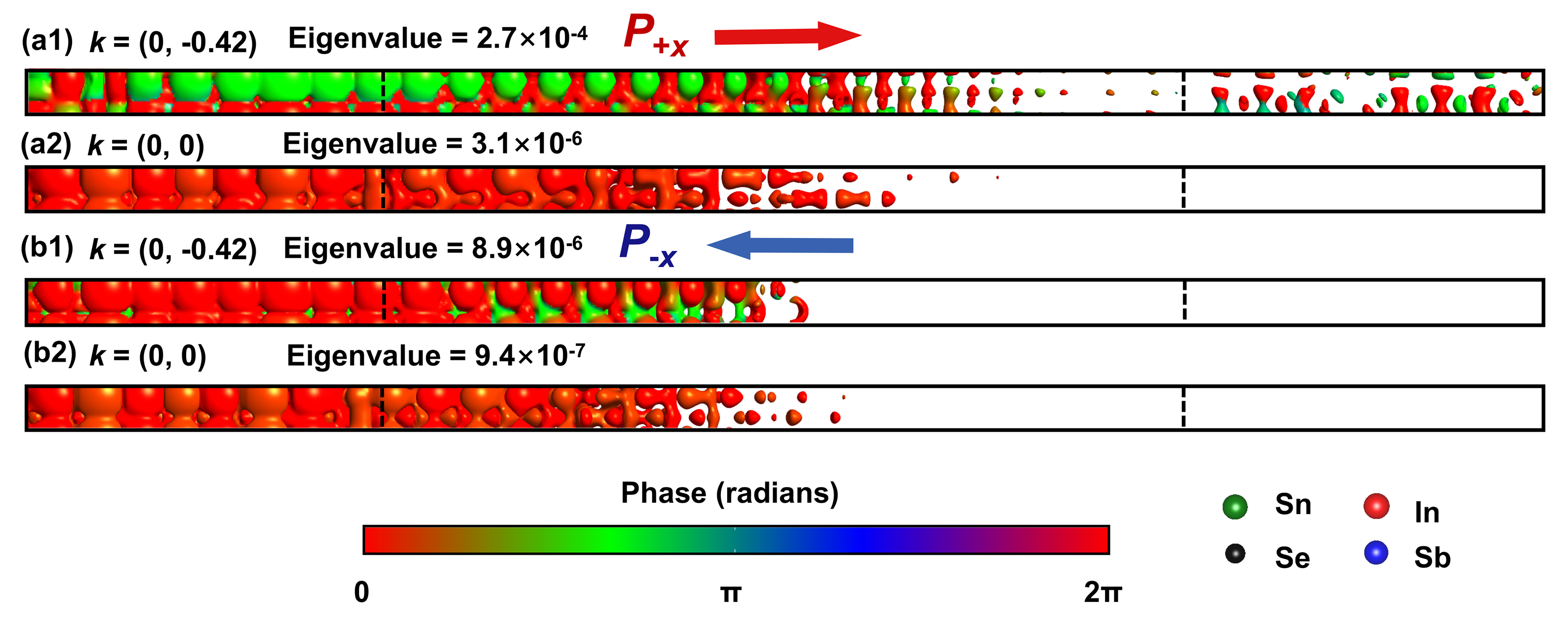}
              \caption{\label{fig:eigenstate} Caption continues on the next page.}
\end{figure}

  Figure~\ref{fig:eigenstate}~shows the transmission eigenstates
  for the two studied $k$-points.
  By comparing Figure~\ref{fig:eigenstate}(a1) with (a2), or
  by comparing Figure~\ref{fig:eigenstate}(b1) with (b2), we can
  find the amplitude of transmission eigenstates
  in the middle to the right of the scattering region
  for $k$ = (0, -0.42) is much
  larger than that for $\Gamma$ point,
  which indicates
  greater transmission probability (this can be read off
  from the eigenvalues shown in Figure~\ref{fig:eigenstate})
  in the case of $k$ = (0, -0.42) in respective
  $P_{+x}$ and $P_{-x}$ states.
  This is particularly
  obvious in $P_{+x}$ state.
  We also find the tranmission eigenstate for $k$ = (0, +0.42)
  is similar to that for $k$ = (0, -0.42), which is not shown
  here.
  The difference of the transmission eigenstates rises from the
  different decay rates of the states at these $k$-points.
  For both $k$ = (0, -0.42) and $\Gamma$ point,
  the transmission eigenstate decays
  through the tunneling barrier. However, the decay rate
  for $k$ = (0, -0.42) is much smaller than that for $\Gamma$ point, \ie,
  the effective transmission channel is suppressed around $\Gamma$ point.
  This explains why the resonant tunneling mainly
  occurs in the region around $k_y$ = +0.42/-0.42, but not the
  $\Gamma$ point.

\addtocounter{figure}{-1}
\begin{figure}[t]
      \caption{
      The transmission eigenstates in the central region at the Fermi level for
      $k$ = (0, -0.42) and $k$ = (0, 0).
      Here, the central region is consisted of the left/right extension layer
      (buffer layer, 2 u.c. doped SnSe)
      and the ferroelectric barrier layer (as thick as 18 u.c. SnSe).
      (a) For $P_{+x}$ state: (a1) $k$ = (0, -0.42); (a2) $k$ = (0, 0).
      (b) For $P_{-x}$ state: (b1) $k$ = (0, -0.42); (b2) $k$ = (0, 0).
      In all the panels, we use same isovalue for comparison.
      The color bar indicates the phase of the eigenstate.
      The vertical dashed black lines denote the
      interfaces between the ferroelectric SnSe and the buffer layers.
      }
  \end{figure}


\section{The width of extension layer in transport calculations}
\begin{figure}[!htp]
	\includegraphics[angle=0,width=0.8\textwidth]{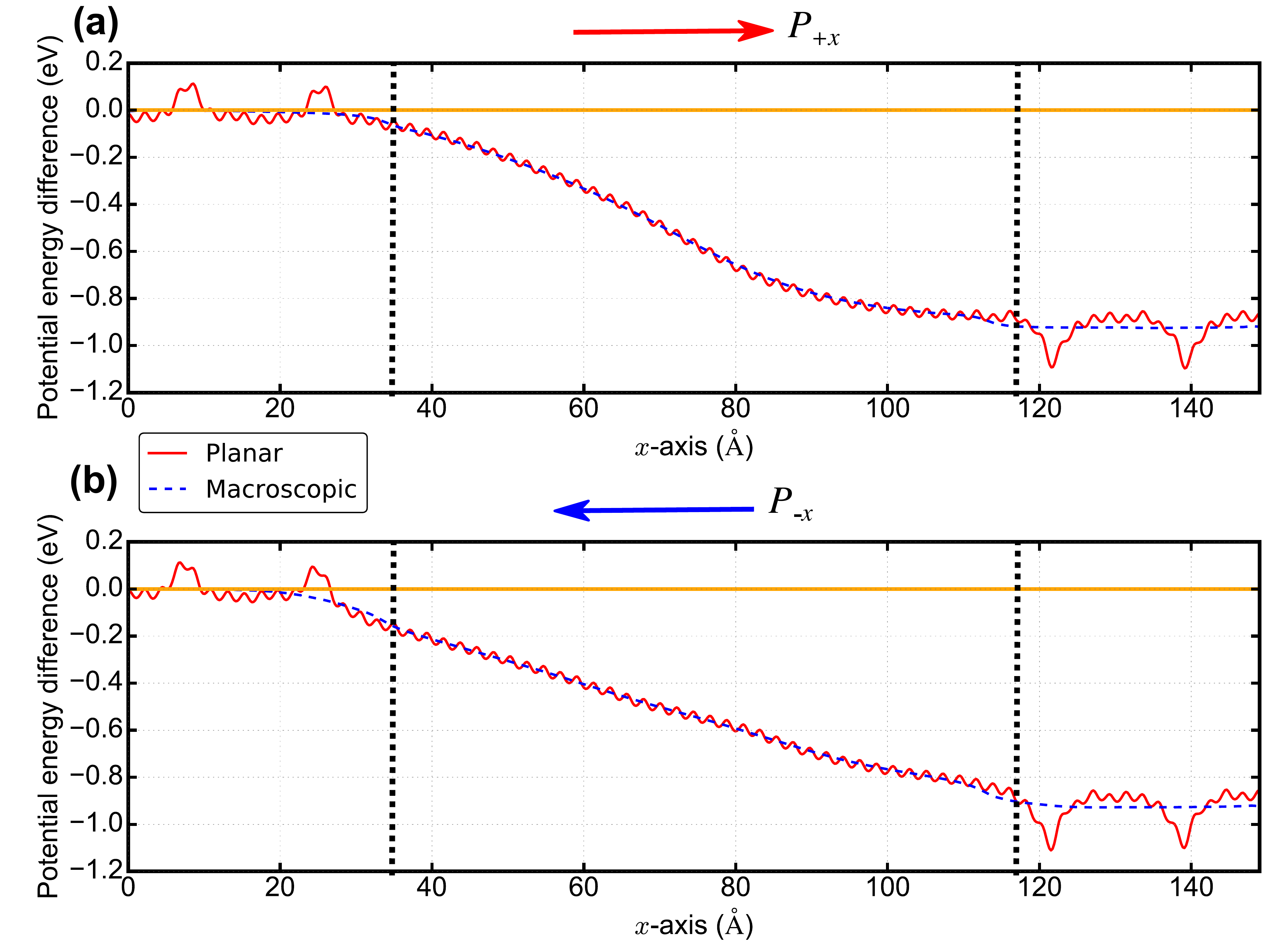}
            \caption{\label{fig:potential-difference-atk} Caption continues on the next page.}
\end{figure}

 In DFT+NEGF calculations,
 the extension layer (buffer layer)
  should be large enough to screen the electrostatic potential.
In our study, the left/right extension layer is as wide as
around 35 \AA, which is almost equivalent to the width of 8 u.c. SnSe.
  Figure~\ref{fig:potential-difference-atk} shows the
planar average (solid red curve)
and macroscopic average (dashed blue curve) electrostatic potential
energy differences obtained in our DFT+NEGF calculations.
We can find that the macroscopic average potential differences are nearly flat
for both left and right polarization states, which indicates our extension layer
is large enough to screen the electrostatic potential.

\addtocounter{figure}{-1}
\begin{figure}[t]
\caption{
The electrostatic potential energy differences
between the electrostatic potential of the
self-consistent valence charge density
(the solution to the Poisson equation with this density)
and the electrostatic potential from a superposition of atomic valence densities
for
(a) $P_{+x}$ and (b) $P_{-x}$ states obtained in our DFT+NEGF calculations.
The solid red curve and dashed blue curve refer to planar average
and macroscopic average potential energy differences, respectively.
The vertical dashed black lines denote the interfaces between
the ferroelectric SnSe and the buffer layers.
}
\end{figure}


\section{Hubbard $U$ effect on the TER effect}

\begin{figure}[!htp]
	\includegraphics[angle=0,width=0.6\textwidth]{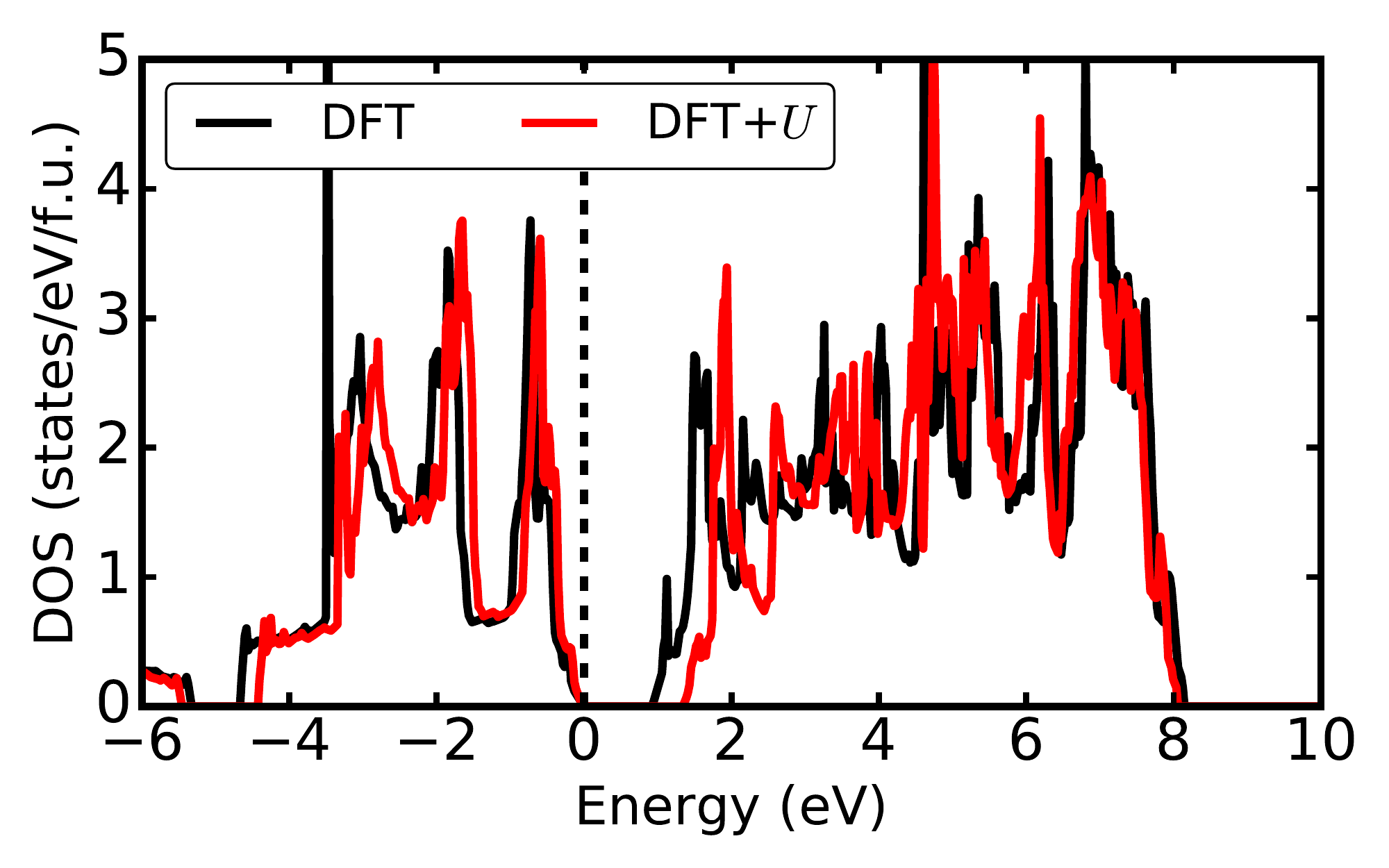}
\caption{\label{fig:DOS-U}
The total density of states of SnSe which is normalized to per formula unit.
The black and red curves refer to DFT and DFT+$U$ ($U$ = 4 eV) calculated DOS, respectively.
The dashed vertical line indicate the Fermi level.
}
\end{figure}

  In Ref.~\citen{Babuka-RSCadv2017}, Hubbard $U$ = 4 eV is used to reproduce the band gaps of Sn-based ferroelectric semiconductors,
  the reported band gaps are in good agreement with the exprimental values.
  Here, in order to have a better estimation of the band gap of SnSe, we also introduce Hubbard $U$ = 4 eV into
  our DFT and DFT+NEGF calculations. The Hubbard $U$ is applied to the $p$ orbitals of Sn.
  Figure~\ref{fig:DOS-U} shows the comparison of band gaps of SnSe between DFT and DFT+$U$ calculations.
  In DFT and DFT+$U$ calculations, the band gaps of SnSe are 0.9 and 1.4 eV, respectively.
  Compared to the experimental band gap ($\sim$ 1.6 eV) of monolayer SnSe~\cite{ShiAdvSci-SnSe}, DFT+$U$ study
  shows better estimation.
  We thus revisit the electron transport properties of 2D-FTJ In:SnSe/SnSe/Sb:SnSe homostructure by
  DFT+$U$+NEGF ($U$ = 4 eV)calculations. The calculated TER is $\sim$ 1120$\%$.
  In our main text, the  DFT+NEGF calculated TER is $\sim$ 1460$\%$. We can find the order magnitude is not changed
  by the included Hubbard $U$. Hence, we think the TER of our studied 2D-FTJ is weakly affected by the
  estimation of band gap of SnSe.
  In addition, we note that the correction of band gap of the ferroelectric barrier material
in the conventional FTJ \SRO/\PTO/Pt junction reported by Tao~\etal~\cite{TaoLL-JAP2016} is also found to
have minor effect on the main results of tunneling properties.
  This implies DFT+NEGF calculations can give very good estimation of the transport properties in FTJs.

\bibliography{2D_FTJ}